\documentclass[12pt]{iopart}

\usepackage{citesort}
\usepackage{iopams}
\usepackage{graphicx}
\usepackage{dcolumn}

\renewcommand{\Im}{\mathrm{Im}}
\renewcommand{\Re}{\mathrm{Re}}

\newcommand{\p}{\partial}

\begin{document}

\title[Stability of lambda and tripod systems]{Stability of linear and non-linear lambda and tripod systems in the presence of amplitude damping}




\author{Viktoras Pyragas$^{1,2}$ and Gediminas Juzeli\={u}nas$^1$}
%
\address{$^1$ Institute of Theoretical Physics and Astronomy of
Vilnius University, LT-01108 Vilnius, Lithuania}
\address{$^2$ Semiconductor Physics Institute of Center for Physical
Sciences and Technology, LT-01108 Vilnius, Lithuania}
\eads{\mailto{viktpy@pfi.lt}, \mailto{gediminas.juzeliunas@tfai.vu.lt}}

\date{\today}

\begin{abstract}
We present the stability analysis of the dark states in the
adiabatic passage for the linear and non-linear lambda and tripod
systems in the presence of amplitude damping (losses). We perform
an analytic evaluation of the real parts of eigenvalues of the
corresponding Jacobians, the non-zero eigenvalues of which are
found from the quadratic characteristic equations, as well as by
the corresponding numerical simulations. For non-linear systems,
we evaluate the Jacobians at the dark states. Similarly to the
linear systems, here we also find the non-zero eigenvalues from
the characteristic quadratic equations. We reveal a common
property of all the considered systems showing that the evolution
of the real parts of eigenvalues can be split into three stages.
In each of them the evolution of the stimulated Raman adiabatic
passage (STIRAP) is characterized by different effective
dimension. This results in a possible adiabatic reduction of one
or two degrees of freedom.

\end{abstract}



\submitto{jpb}

\maketitle

\section{\label{sec:1}Introduction}

Over the last couple of decades there has been a continuing
interest in the stimulated Raman adiabatic passage (STIRAP)
\cite{berg98,vitan01,vitan01a,arim96,shap07}. The simplest
situation is the adiabatic passage in a linear lambda system
\cite{oreg84,carr88,vitan97a,unan97,vitan04,vitan05,vitan09}
containing a single dark (uncoupled) state which is immune to the
atom-light coupling. If atomic initial and final states are the
ground states representing the dark states of the system, the atom
can be transferred between these two states by slowly changing the
relative intensity of the laser pulses. When the adiabatic passage
is slow enough, the excited state is only slightly populated and
thus the losses are minimum. The analysis has been extended for
the STIRAP process in the tripod system characterized by two dark
states  \cite{unan98,unan99,vitan10}. This enables to create a
quantum superposition of metastable states out of a single initial
state in a robust and coherent way \cite{unan99a,goto07}. The
schemes involving more atomic and molecular levels were also
proposed for creation a superposition of states \cite{kis05} as
well as for experimental control of excitation flow
\cite{ekers06}. Recently the treatment was further extended to the
non-linear lambda
\cite{wink05,moal06,ling04,pu07,ling07,itin07,itin08,hope01,meng08,mackie00,cheng06,zhao08}
and tripod \cite{zhou10} schemes.

Usually the STIRAP is based on the adiabatic approximation.
However, one has to distinguish between the adiabatic
approximation and the adiabatic reduction of dynamic systems. In
quantum mechanics, a closed quantum system is said to undergo
adiabatic dynamics if its Hilbert space can be decomposed into
decoupled Schr\"{o}dinger eigenspaces with distinct,
time-continuous, and non-crossing instantaneous eigenvalues of
Hamiltonian. On the other hand an open quantum system is said to
undergo adiabatic dynamics if its Hilbert-Schmidt space can be
decomposed into decoupled Lindblad-Jordan eigenspaces with
distinct, time-continuous, and non-crossing instantaneous
eigenvalues of the Lindblad superoperator \cite{lidar05}. The
system is called adiabatically approximated if the error term in
the Schr\"{o}dinger equation (for closed systems) or in the master
equation (for open systems) is much less than the diagonalysed
part; i.e. one may neglect the non-diagonal terms in order to get
an adiabatically approximated version of the system. Note that in
the presence of fast driven oscillations some additional
conditions (in addition to the slowness of the evolution of the
Hamiltonian) have to be imposed \cite{amin09}.

Another procedure is the adiabatic elimination of decaying degrees
of freedom. It is related to the dynamic systems in which some
degrees of freedom may decay. In accordance with this definition,
these degrees of freedom may be adiabatically eliminated by
solving the corresponding algebraic equations; the r.h.s. of
decaying equations are set to be equal to zero. Consequently, one
obtains the dependencies of decaying variables on the remaining
ones. If there are several decaying variables, one may eliminate
them one by one, starting from the fastest variable, and finishing
with the slowest one. Such a procedure can be found e.g. in the
book by Haken \cite{haken83}.

The aim of the present work is to perform a stability analysis of
the dark states in the adiabatic passage for the linear and
non-linear lambda and tripod systems. The analysis sets limits to
the adiabatic reduction of the systems. Moreover, we have revealed
that in all the considered systems, the stability properties of
the dark states are similar, namely, there are three time
intervals with different number of the negative real parts of the
Jacobians. This suggests that the corresponding linear and
non-linear systems have equal possibilities of adiabatic
reduction. Although the linear lambda
\cite{berg98,vitan01,vitan01a,arim96,shap07,oreg84,carr88,vitan97a,unan97,vitan04,vitan05,vitan09}
and tripod systems
\cite{unan98,unan99,vitan10,unan99a,goto07,wu07} have been
substantially studied in the literature, here we apply our
treatment also to these linear setups in order to facilitate the
subsequent analysis of the non-linear systems.

The non-linear lambda system can be realized in the Bose-Einstein
condensates (BEC) via photoassociation (PA) from a dissociated
(quasicontinuum) atomic state to the ground molecular state in the
presence of the intermediate molecular state \cite{wink05,moal06}.
The aim of the quantum control is to transfer the whole population
from the dissociated atomic to the ground molecular state. One
thus creates ultracold molecules by associating cold atoms
\cite{mackie00,mackie05}. In this case the dark state is a
generalisation of that for the linear systems. In the linear case,
the dark state is defined as a superposition of the initial and
target ground atomic states which corresponds to the zero
eigenergy of the system Hamiltonian. The same dark state may be
also defined as a steady state solution of the Schr\"{o}dinger
equation. If we consider the nonlinear system, we can again define
the dark state as a steady state \cite{mackie00} of the equations
of motion following from the Heisenberg equation. In the nonlinear
systems, the behaviour of the dark state reproduces that in the
linear systems although the superposition is missing now. We thus
get the nonlinear version of STIRAP: the entire population is
distributed among the steady state probability amplitudes of the
initial atomic and the target molecular state. At the beginning
the whole population is atomic, whereas at the end it is in the
ground molecular state.

Differently from the traditional STIRAP in an atomic lambda
system, the atom-molecule STIRAP contains nonlinearities
originating from the conversion process of atoms to molecules, as
well as from the interparticle interactions described by the
non-linear mean-field contributions. The existence of such
nonlinearities makes it difficult to analyse the adiabaticity of
the atom-molecule conversion systems due to the absence of the
superposition principle. In the STIRAP, the linear instability
could make the quantum evolution deviate from the dark state
rapidly even in adiabatic limit \cite{ling04}. Therefore, it is
important to avoid such an instability for the efficiency of the
STIRAP.

The non-linear tripod system can be realized in the PA with two
target states involved. Specifically, one may consider the
atom-molecule transition in ultracold quantum gases via PA. It was
first considered in \cite{zhou10} where the second order dynamic
system was derived that parametrizes the solution evolving on the
dark state manifold. However the stability of the solution moving
along the manifold was not considered. Therefore we shall check
the stability of this solution, i.e. see whether the nearby
solutions are attracted back to this manifold.

The adiabatic theory for non-linear quantum systems was first
discussed by Liu {\it et al.} \cite{liu03} who obtained the
adiabatic conditions and adiabatic invariants by representing the
non-linear Heisenberg equation in terms of an effective classical
Hamiltonian. Pu {\it et al.} \cite{pu07} and Ling {\it et al.}
\cite{ling07} extended such an adiabatic theory to the atom-dimer
conversion system by linking the nonadiabaticity with the
population growth in the collective excitations of the dark state.
Specifically, it was shown that a passage is adiabatic if the
solution remains in a close proximity to the dark state.  Itin and
Watanabe \cite{itin07} presented an improved adiabatic condition
by applying methods of the classical Hamiltonian dynamics. The
atom-molecule dark-state technique in the STIRAP was theoretically
generalized to create more complex homonuclear or heteronuclear
molecule trimers or tetramers
\cite{jing07,jing08a,jing08b,jing08c}.

An important issue is the instability and the adiabatic property
of the dark state in such complex systems. For example, the
dynamics of a non-linear lambda system describing BEC of atoms and
diatomic molecules was studied and a model of the dark state with
collisional interactions was investigated \cite{itin08,hope01}. It
was shown that non-linear instabilities can be used for precise
determination of the scattering lengths. On the other hand, the
transfer of atoms to molecules via STIRAP is robust with respect
to detunings, $\chi^{3}$ nonlinearities, and small asymmetries
between the peak strengths of the two Raman lasers \cite{hope01}.
The complete conversion is destroyed by spatial effects unless the
time scale of the coupling is much faster than the pulse duration.
In addition, a set of robust and efficient techniques  has been
introduced \cite{lew04} to coherently manipulate and transport
neutral atoms based on three-level atom optics (TLAO).

It is to be noted that the dynamics of an adiabatic sweep through
a Feshbach resonance was studied \cite{pazy05} in a quantum gas of
fermionic atoms. An interesting application of BEC is an atom
diode with a directed motion of atoms \cite{larson11}. Another
example of BEC was presented in \cite{meystre11} where it was
shown that the two-colour PA of fermionic atoms into bosonic
molecules via a dark-state transition results in a significant
reduction of the group velocity of the photoassociation field.
This is similar to the Electromagnetically induced transparency
(EIT) in atomic systems characterized by the three-levels of the
lambda type. In addition the coupled non-linear Schr\"{o}dinger
equations have been considered \cite{zhang11} to describe the
atomic BECs interacting with the molecular condensates through the
STIRAP loaded in an external potential. The results have shown
that there is a class of external potentials where the exact dark
solutions can be formed. In \cite{kis02} it was shown that it is
possible to perform qubit rotations by STIRAP, and proposed a
rotation procedure in which the resulting state corresponds to a
rotation of the qubit, with the axis and angle of rotation
determined uniquely by the parameters of the laser fields.

A relevant tool for studying the adiabaticity is the adiabatic
fidelity. It indicates how far is the current solution of the
system from the dark state. Meng {\it et al.} have generalized the
definition of fidelity for the non-linear system \cite{meng08}.
They have studied the dynamics and adiabaticity of the population
transfer for atom-molecule three-level lambda system on a STIRAP.
It was also discussed how to achieve higher adiabatic fidelity for
the dark state through optimizing the external parameters of the
STIRAP. In the subsequent paper \cite{meng09} Meng {\it et al.}
have used the same definition of adiabatic fidelity in order to
discuss the adiabatic evolution of the dark state in a non-linear
atom-trimer conversion due to a STIRAP. It is to be noted that
Ivanov and Vitanov have recently proposed novel high-fidelity
composite pulses for the most important single-qubit operations
\cite{vitan11}.

In this work, we analyse the problem of reducing the dimension
(simplifying) in the linear/non-linear three- and four-level
models. This procedure is called the adiabatic reduction, and its
validity is closely related with the theory of adiabaticity
discussed above. The exact three- or four-level system may be
adiabatically reduced to a system with lower dimension. The
question that arises is how many dynamic variables can be
eliminated? In other words, what is the effective dimension of the
system? The answer lies in the eigenvalues of the Jacobian
computed at the dark state. The zero real parts of eigenvalues
mean that in some directions the nearby solutions are behaving
neutrally in respect to the dark state. The negative real parts in
turn mean that some directions are stable, and the nearby
solutions converge towards the dark state. Therefore we conclude
that the number of negative real parts dictates the number of
variables that can be adiabatically eliminated (see e.g.
\cite{haken83}). On the other hand, the number of zero real parts
yields the effective dimension of the system. Note that we find
the non-zero eigenvalues analytically from quadratic
characteristic equations.

One of the central issues in our work is the presence of
dissipation in all the considered systems. The non-zero losses
make the adiabatic reduction easier to implement since the term of
losses acts as a "controller" that attracts the nearby solutions
towards the dark state. However, Vitanov and Stenholm have
demonstrated that the losses cause also the decrease of transfer
efficiency to the target state \cite{vitan97a}. This decrease can
be circumvented by higher pulse areas since the range of decay
rates over which the transfer efficiency remains high, has been
found to be proportional to the squared pulse area (see (10) in
\cite{vitan97a}). In the subsequent developments the effect of
spontaneous emission on the population transfer efficiency in
STIRAP was explored for the linear lambda \cite{vitan04,vitan05}
and tripod  \cite{vitan10} systems. In addition, Renzoni {\it et
al.} \cite{renzoni97} have considered the
coherent-population-trapping (CPT) phenomenon in a thermal sodium
atomic beam. It was demonstrated that CPT may be realized on those
open transitions with an efficiency decreasing with the amount of
spontaneous emission towards external states. On the other hand,
here we concentrate on the stability issues of the linear and
non-linear lambda and tripod systems with the losses.

The paper is organized as follows. In next two sections we shall consider the stability of the
linear lambda and tripod systems with losses. In sections
\ref{sec:4} and \ref{sec:5} the analysis is extended to the non-linear lambda and tripod systems.
In section \ref{sec:6} we discuss the role of the one-photon detuning followed by the conclusions in section \ref{sec:7}.

\section{\label{sec:2}The linear lambda system}

\begin{figure}[h!]
\centering\includegraphics[width=0.45\textwidth]{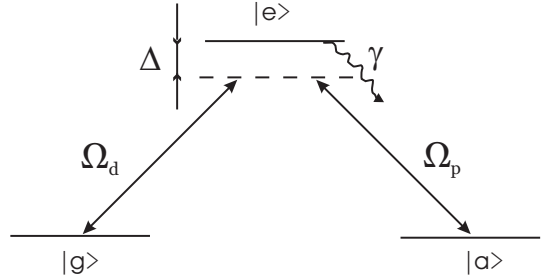}
\caption{\label{Fig_L_scheme} Three-level system coupled by two
lasers. $\Omega_{\rm{p}}$ and $\Omega_{\rmd}$ are the Rabi frequencies for the
pump and damp laser, $\Delta$ is the one photon detuning, and
$\gamma$ is the loss rate.}
\end{figure}

In this section we shall provide a summary on the STIRAP in the
linear lambda system with losses studied in
\cite{vitan97a,vitan04,vitan05} followed by the stability analysis
of the system. The three-level lambda system is shown in figure
\ref{Fig_L_scheme}. The excited state $|\rme\rangle$ is coupled to
two ground states $|\rm{a}\rangle$ and $|\rm{g}\rangle$ with the coupling
strengths denoted as $\Omega_{\rm{p}}$ and $\Omega_{\rmd}$ to form a
lambda scheme. The Hamiltonian of such system reads:
\begin{equation}
\label{ties L ham}
\begin{array}{l}
    H = -\hbar(\Delta + \rmi\gamma)|\rme\rangle\langle \rme| + \frac{\hbar}{2} [\Omega_{\rm{p}}|\rm{a}\rangle\langle \rme| + \Omega_{\rm{d}}|\rm{g}\rangle\langle \rme| + H.c.].
\end{array}
\end{equation}

Note that in this Hamiltonian besides the real-valued one-photon
detuning $\Delta$ there is the imaginary term $\gamma$
representing the losses. Denoting the amplitudes of the
three-level state as $\psi_{\rme}$, $\psi_{\rm{a}}$, and
$\psi_{\rm{g}}$ respectively, we get the Schr\"{o}dinger equation:
\begin{eqnarray}
    \rmi\dot{\psi}_{\rm{a}} &=& \Omega_{\rm{p}}\psi_{\rme},\label{L_lin_dina}\\
    \rmi\dot{\psi}_{\rme} &=& -(\Delta + \rmi\gamma)\psi_{\rme} +
    \Omega_{\rm{p}}\psi_{\rm{a}} +
    \Omega_{\rm{d}}\psi_{\rm{g}}, \label{L_lin_dine}\\
    \rmi\dot{\psi}_{\rm{g}} &=& \Omega_{\rm{d}}\psi_{\rm{e}}.\label{L_lin_ding}
\end{eqnarray}
The normalization reads:
\begin{equation}\label{L_lin_norm}
    |\psi_{\rm{a}}(t)|^{2} + |\psi_{\rm{g}}(t)|^2 + |\psi_{\rme}(t)|^2 \leq 1,
\end{equation}
where equality holds for initial time. Because of losses ($\gamma
> 0$) the total normalization will be slightly reduced (for $t>0$)
during the transfer of population through the excited level. This
property of the total population is assumed throughout the paper.

We take the laser pulses to be Gaussian
\begin{eqnarray}
   \Omega_{\rmd}  &=& \Omega_{0} \exp\left[-\frac{(t-t_{\rmd})^2}{T^{2}}\right],\label{L_lin_Omd}\\
   \Omega_{\rm{p}}  &=& \Omega_{0} \exp\left[-\frac{(t-t_{\rm{p}})^2}{T^{2}}\right].\label{L_lin_Omp}
\end{eqnarray}
Here the pulses are centered at $t_{d}$ and $t_{p}$, respectively,
$T$ is their width, and $\Omega_{0}$ is their amplitude.

The third order system (\ref{L_lin_dina},\ref{L_lin_dine},\ref{L_lin_ding}) may be rewritten in a matrix form
\begin{equation}\label{L_din_Amatr}
    \dot{\bPsi} = -\rmi H\bPsi \equiv A\bPsi,
\end{equation}
where $\bPsi=[\psi_{\rm{a}},\psi_{\rme},\psi_{\rm{g}}]^T$ is the vector of the
state of the system, and


\begin{equation}\label{L_din_H}
    H = \left(\begin{array}{ccc}
      0 & \Omega_{\rm{p}} & 0 \\
      \Omega_{\rm{p}} & -(\Delta+\rmi\gamma) & \Omega_{\rm{d}} \\
      0 & \Omega_{\rm{d}} & 0
    \end{array} \right)
\end{equation}
is the corresponding Hamiltonian. The matrix $A \equiv -\rmi H$ is
the Jacobian of the system. If the Hamiltonian possesses
eigenvalues $\omega$, then the eigenvalues of the Jacobian are
defined by $\lambda=-\rmi\omega$. Note that the real parts of
$\lambda$ determine the stability of the fixed point at the
origin.

We now find these eigenvalues, more specifically their real parts.
The eigenvalues of Hamiltonian satisfy the characteristic equation
\begin{equation}\label{L_detH}
    \det||H-I\omega||=0,
\end{equation}
with $I$ denoting the unit matrix. Expanding the corresponding
third-order determinant one finds that one eigenvalue is always
zero:
\begin{equation}\label{omega1}
    \omega_1=0.
\end{equation}
The other two eigenvalues satisfy the quadratic equation:
\begin{equation}\label{omega_qeq}
    \omega^2+(\Delta+\rmi\gamma)\omega-(\Omega_{\rm{d}}^2+\Omega_{\rm{p}}^2)=0.
\end{equation}
Thus the two eigenvalues satisfy
\begin{eqnarray}
\omega_2+\omega_3 &=& -[\Delta+\rmi\gamma],\label{L_om23a}\\
\omega_2 \omega_3 &=& -[\Omega_{\rm{d}}^2+\Omega_{\rm{p}}^2].\label{L_om23b}
\end{eqnarray}
Assuming $\Delta=0$, the solutions of
(\ref{omega_qeq}) read:
\begin{equation}\label{L_om23_sol}
    \omega_{2,3} = \{-\rmi\gamma \pm [-\gamma^2 + 4(\Omega_{\rmd}^2+\Omega_{\rm{p}}^2)]^{1/2}\}/2.
\end{equation}
For $t \rightarrow \pm \infty$ the Rabi frequencies go to zero,
and it follows from (\ref{L_om23a},\ref{L_om23b}) that $\omega_2=0, \omega_3 =
-\rmi\gamma$. Calling on (\ref{omega1}), we can write
\begin{equation}\label{L_lt_infty}
    \lambda_{1,2}=0, \quad \lambda_3=-\gamma,
\end{equation}
or $\Re(\lambda_{1,2})=0$, and $\Re(\lambda_3)=-\gamma$
for $t \rightarrow \pm \infty$.

For finite times there is a region where the Rabi frequencies are
large enough, so that the discriminant is positive  in
(\ref{L_om23_sol}): $D\equiv -\gamma^2 +
4(\Omega_{\rmd}^2+\Omega_{\rm{p}}^2)>0$. Such a situation occurs in a
certain interval $t_1<t<t_2$, and from (\ref{L_om23_sol}) we get
\begin{eqnarray}
\omega_{2,3} &=& -\rmi\gamma/2 \pm \sqrt{D}/2,\label{L_roots23a}\\
\lambda_{2,3} &=& -\gamma/2 \mp \rmi\sqrt{D}/2.\label{L_roots23b}
\end{eqnarray}
The first eigenvalue is $\lambda_1=\omega_1=0$. The boundaries
$t_1,t_2$ are solutions of $D(t)=0$ in respect to time.

Hence, in the interval $t_1<t<t_2$, the real parts are $\Re(\lambda_1)=0$,
$\Re(\lambda_{2,3})=-\gamma/2$.

We may adiabatically reduce the dimension of this system but we
first transform its variables. We change the bare variables
$\psi_{\rm{a}}$, $\psi_{\rm{g}}$ to the bright $\psi_{\rm{B}}$ and dark $\psi_{\rm{D}}$ one:
\begin{eqnarray}
    \psi_{\rm{B}} &=& (\Omega_{\rm{p}}\psi_{\rm{a}} + \Omega_{\rmd}\psi_{\rm{g}})/\Omega,\label{varch_LlB}\\
    \psi_{\rm{D}} &=& (\Omega_{\rmd}\psi_{\rm{a}} - \Omega_{\rm{p}}\psi_{\rm{g}})/\Omega,\label{varch_LlD}
\end{eqnarray}
where
\begin{equation}\label{L_lin_Om}
    \Omega = (\Omega_{\rm{p}}^2 + \Omega_{\rmd}^2)^{1/2}.
\end{equation}

Denoting $\xi_{\rm{p}} = \Omega_{\rm{p}}/\Omega$, $\xi_{\rmd} = \Omega_{\rmd}/\Omega$, and
performing some operations, we obtain the following equations for
the new variables:
\begin{eqnarray}
    \rmi\dot{\psi}_{\rm{B}} &=& \alpha\psi_{\rm{D}} + T\Omega\psi_{\rme},\label{L_lin_varch_dinB}\\
    \rmi\dot{\psi}_{\rme} &=& -T(\Delta + \rmi\gamma)\psi_{\rme} +
    T\Omega\psi_{\rm{B}}, \label{L_lin_varch_dine}\\
    \rmi\dot{\psi}_{\rm{D}} &=& \alpha^{*}\psi_{\rm{B}},\label{L_lin_varch_dinD}
\end{eqnarray}
where we have made the time dimensionless by substituting $t/T
\rightarrow t$. Here $\alpha=\rmi
(\dot{\xi}_{\rm{p}}\xi_{\rmd}-\dot{\xi}_{\rmd}\xi_{\rm{p}})$ is a
dimensionless parameter in which derivatives
$\dot{\xi}_{\rm{p}},\dot{\xi}_{\rm{d}}$ are taken in respect to
dimensionless time.

We now adiabatically eliminate the amplitude $\psi_{\rme}$ by
setting $\dot{\psi}_{\rme} = 0$. From (\ref{L_lin_varch_dine}) we
get
\begin{equation}\label{L_lin_ad_psie}
    \psi_{\rme} = \frac{\Omega}{\Delta+\rmi\gamma}\psi_{\rm{B}}.
\end{equation}
Inserting this result in (\ref{L_lin_varch_dinB},\ref{L_lin_varch_dinD}) we obtain
\begin{eqnarray}
    \rmi\dot{\psi}_{\rm{B}} &=& \alpha\psi_{\rm{D}} + \frac{T\Omega^2}{\Delta+\rmi\gamma}\psi_{\rm{B}},\label{L_lin_varch_adB}\\
    \rmi\dot{\psi}_{\rm{D}} &=&
    \alpha^{*}\psi_{\rm{B}}.\label{L_lin_varch_adD}
\end{eqnarray}
We solve this system to find the dynamics of $\psi_{\rm{B}}$,
$\psi_{\rm{D}}$, and from (\ref{L_lin_ad_psie}) we find the dynamics of
$\psi_{\rme}$.

The system (\ref{L_lin_varch_adB},\ref{L_lin_varch_adD}) may be also adiabatically
reduced. We now set $\dot{\psi}_{\rm{B}} = 0$, and solve
(\ref{L_lin_varch_adB}) for $\psi_{\rm{B}}$:
\begin{equation}\label{L_lin_ad2_psiB}
    \psi_{\rm{B}} =
    -\frac{\alpha(\Delta+\rmi\gamma)}{T\Omega^2}\psi_{\rm{D}}.
\end{equation}
Inserting this result in (\ref{L_lin_varch_adD}), we get a first
order dynamic system
\begin{equation}\label{L_din_lin_varch_ad2}
    \rmi\dot{\psi}_{\rm{D}} =
    -\frac{|\alpha|^2(\Delta+\rmi\gamma)}{T\Omega^2}\psi_{\rm{D}}.
\end{equation}

The Hamiltonian for the $2$D reduced system
(\ref{L_lin_varch_adB},\ref{L_lin_varch_adD}) reads:
\begin{equation}\label{L_din_H2}
    H_{2\rm{D}} = \left(\begin{array}{cc}
      \frac{T\Omega^{2}}{\Delta+\rmi\gamma} & \alpha \\
      \alpha^{*} & 0
    \end{array} \right)
\end{equation}
Solving the characteristic equation for this Hamiltonian, and
assuming $\Delta=0$, one obtains two eigenvalues:
\begin{eqnarray}
\omega_{1,2} &=& \frac{\rmi}{2}\left[-\frac{T\Omega^{2}}{\gamma} \pm \left(\frac{T^{2}\Omega^{4}}{\gamma^{2}}-4|\alpha|^{2}\right)^{1/2}\right],\label{L_roots2Da}\\
\lambda_{1,2} &=& \frac{1}{2}\left[-\frac{T\Omega^{2}}{\gamma} \pm
\left(\frac{T^{2}\Omega^{4}}{\gamma^{2}}-4|\alpha|^{2}\right)^{1/2}\right],\label{L_roots2Db}
\end{eqnarray}
Here $\lambda_{1,2}$ are the eigenvalues of the corresponding
Jacobian defined as $A_{2\rm{D}}\equiv -\rmi H_{2\rm{D}}$. There
are two regimes of evolution of the eigenvalues corresponding to
$D_{2}<0$ and $D_{2}>0$, where
$D_{2}=\frac{T^{2}\Omega^{4}}{\gamma^{2}}-4|\alpha|^{2}$ is the
discriminant in (\ref{L_roots2Da},\ref{L_roots2Db}).

Hence, we have the eigenvalues for all the three versions of the
linear lambda system. For the exact $3$D system, they are given by
$\lambda_{1}=\omega_{1}=0$ and
(\ref{L_roots23a},\ref{L_roots23b}). For the reduced $2$D system,
they read (\ref{L_roots2Da},\ref{L_roots2Db}). Finally, a trivial
single eigenvalue for the $1$D system follows from
(\ref{L_din_lin_varch_ad2}):
\begin{equation}\label{L_eigv_ad1}
    \lambda_{1} = -\rmi \omega_{1} =
    -\frac{|\alpha|^2\gamma}{T \Omega^2}.
\end{equation}

\begin{figure}[h!]
\centering\includegraphics[width=0.45\textwidth]{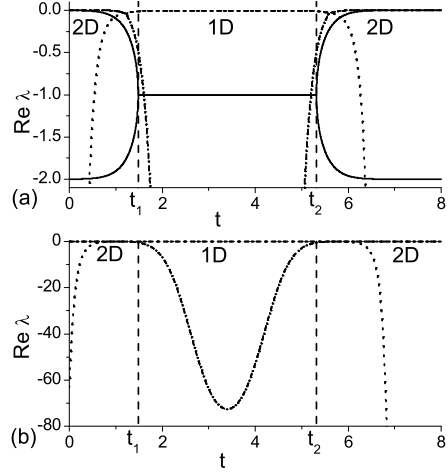}
\caption{\label{Fig_L_Aties} (a) Dynamics of real parts of
eigenvalues of the Jacobian for linear lambda system, computed for
all three cases ($3$D, $2$D, and $1$D) by
(\ref{L_roots23a},\ref{L_roots23b}),
(\ref{L_roots2Da},\ref{L_roots2Db}) and (\ref{L_eigv_ad1}),
respectively. The dashed vertical lines set the boundaries for the
$1$D and $2$D processes. Here $t_1=1.49$ and $t_2=5.32$. The solid
lines are the real parts for the $3$D, the short dash dotted and
short dash line correspond to the $2$D case, and the dotted line
is the real part for the $1$D reduced system; (b) the dynamics of
real parts for the $2$D and $1$D cases in extended vertical scale.
The lines are chosen in the same way as in (a). The parameters are
as follows: $\Delta = 0$, $\gamma = 2.0$, $\Omega_{0} = 10.0$,
$t_{\rm{p}} = 3.8$, $t_{\rmd} = 3.0$, and $T = 1.0$.}
\end{figure}

In figure \ref{Fig_L_Aties} we show the dynamics of the real parts
of eigenvalues for all three cases.

As the solid lines in figure \ref{Fig_L_Aties}(a) show, for
$t<t_{1}$ and $t>t_{2}$ there are two different non-zero branches
for the $3$D case: the upper branch determines the slow decay of
the bright state, whereas the lower branch causes the fast decay
of the excited one. Therefore, we may eliminate the excited state,
and cannot do this with the bright one. In the middle of the
process, where $t_{1}<t<t_{2}$, both branches become degenerate
with real part of eigenvalues equal to $-\gamma/2$. We thus may
reduce the both states, excited and bright. During the whole
evolution, one real part remains exactly zero showing that the
dark state does not experience any losses. This fact indicates
that the process is adiabatic.

The short dash dotted and short dash lines show the dynamics of
two real parts for the $2$D reduced system. Although the time
moments $[(\tilde{t}_{1},\tilde{t}_{2})=(1.24,5.56)]$ at which the
discriminant $D_{2}$ is zero differ from the times
[$(t_{1},t_{2})=(1.49,5.32)$], where the discriminant $D$ goes to
zero, these differences are small compared to the interval
$(t_{2}-t_{1})$. In the time intervals of $t<t_{1}$ and $t>t_{2}$
the both branches are close to zero indicating that neither bright
nor dark state may be eliminated. The process is here again $2$D.
In contrast, in the range of time, where $t_{1}<t<t_{2}$, the
decay rate of the bright state is very large compared to that of
the dark state. It is well seen in the figure
\ref{Fig_L_Aties}(b). Therefore the process is again $1$D as for
the exact case.

And lastly, the dotted line shows the dynamics of the real part
for the $1$D reduced system. One may clearly see that for
$t<t_{1}$ and $t>t_{2}$ the decay of the dark state is rapid
compared to that for $t_{1}<t<t_{2}$. The fastness of decay for
$t<t_{1}$ and $t>t_{2}$ is seen expressively in the figure
\ref{Fig_L_Aties}(b). This leads to conclusion that the $1$D
reduced system is appropriate only in the time interval of
$t_{1}<t<t_{2}$.

Exactly as for the $3$D system, the evolution of the system in the
$2$D ($1$D) approximations is adiabatic for the whole time of
integration (in the time interval $t_{1}<t<t_{2}$) since for both
cases the decay rate of the dark state is very small compared with
that of the bright state.

\section{\label{sec:3}The linear tripod}

The STIRAP process in the linear tripod scheme without dissipation
was first considered by Unanyan {\it et al.} \cite{unan98,unan99}.
Here we outline this scheme in which the dissipation is also
included. Afterwards we perform the linear analysis of this
system.

\begin{figure}[h!]
\centering\includegraphics[width=0.45\textwidth]{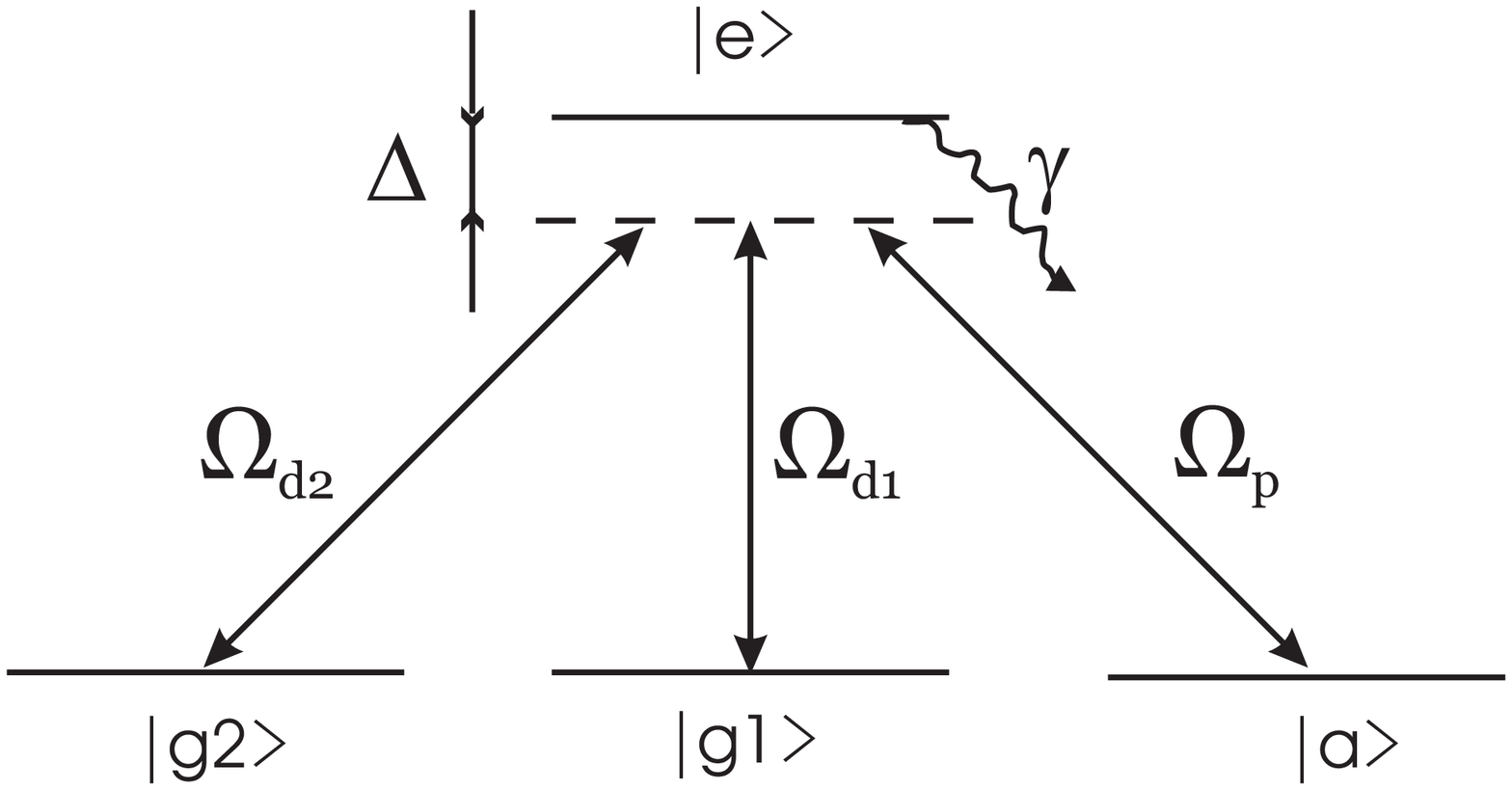}
\caption{\label{Fig_tr_scheme} Four-level system coupled by three
lasers. $\Omega_{\rm{p}}$ and $\Omega_{\rmd 1}$, $\Omega_{\rmd 2}$ are the Rabi
frequencies for the pump, and damp lasers, $\Delta$ is the one
photon detuning, and $\gamma$ is the loss rate.}
\end{figure}

Consider the four-level system schematically shown in figure
\ref{Fig_tr_scheme}. The excited state $|\rme\rangle$ is coupled to
three ground states $|\rm{a}\rangle$, $|\rm{g1}\rangle$, and $|\rm{g2}\rangle$
with the coupling strengths denoted as $\Omega_{\rm{p}}$, $\Omega_{\rmd 1}$,
and $\Omega_{\rmd 2}$, respectively. Here p stands for pump and d
stands for the damp. The four-level Hamiltonian reads:
\begin{equation}
\label{ties tr ham}
\begin{array}{l}
    H = -\hbar(\Delta + \rmi\gamma)|\rme\rangle\langle \rme| + \frac{\hbar}{2} [\Omega_{\rm{p}}|\rm{a}\rangle\langle \rme| + \Omega_{\rmd 1}|\rm{g}1\rangle\langle \rme| + \\
    + \Omega_{\rmd 2}|\rm{g}2\rangle\langle \rme| + H.c.].
\end{array}
\end{equation}

Denoting the amplitudes as $\psi_{\rme}$, $\psi_{\rm{a}}$,
$\psi_{\rm{g}1}$, and $\psi_{\rm{g}2}$, the Schr\"{o}dinger
equation reads:
\begin{eqnarray}
    \rmi\dot{\psi}_{\rm{a}} &=& \Omega_{\rm{p}}\psi_{\rme},\label{trip_lin_dina}\\
    \rmi\dot{\psi}_{\rme} &=& -(\Delta + \rmi\gamma)\psi_{\rme} +
    \Omega_{\rm{p}}\psi_{\rm{a}} +
    \Omega_{\rmd 1}\psi_{\rm{g}1} +\nonumber\\
    &+& \Omega_{\rmd 2}\psi_{\rm{g}2}, \label{trip_lin_dine}\\
    \rmi\dot{\psi}_{\rm{g}1} &=& \Omega_{\rmd 1}\psi_{\rme},\label{trip_lin_ding1}\\
    \rmi\dot{\psi}_{\rm{g}2} &=& \Omega_{\rmd 2}\psi_{\rme}.\label{trip_lin_ding2}
\end{eqnarray}
The normalization is given by
\begin{equation}\label{trip_lin_norm}
    |\psi_{\rm{a}}(t)|^{2} + |\psi_{\rm{g}1}(t)|^2 + |\psi_{\rm{g}2}(t)|^2 + |\psi_{\rme}(t)|^2 \le 1,
\end{equation}
where equality holds for initial time.

The Rabi frequencies are given by
\begin{eqnarray}
   \Omega_{\rm{p}}  &=&\Omega_{0} \exp\left[-\frac{(t-t_{\rm{p}})^2}{T^{2}}\right],\label{trip_lin_Omp}\\
   \Omega_{\rmd 1}  &=& K_{1}\Omega_{0} \exp\left[-\frac{(t-t_{\rmd 1})^2}{T^{2}}\right],\label{trip_lin_Omd1}\\
   \Omega_{\rmd 2}  &=& K_{2}\Omega_{0} \exp\left[-\frac{(t-t_{\rmd 2})^2}{T^{2}}\right].\label{trip_lin_Omd2}
\end{eqnarray}
Here the pulses are centered at $t_{\rm{p}}$, $t_{\rmd 1}$ and
$t_{\rmd 2}$, respectively. $T$ is the width of the pulses,
$K_{1}$, $K_{2}$ determine the amplitudes for the damp pulses, and
$\Omega_{0}$ is the amplitude for the pump pulse.

The system (\ref{trip_lin_dina},\ref{trip_lin_dine},\ref{trip_lin_ding1},\ref{trip_lin_ding2}) can be written in the form of
(\ref{L_din_Amatr}) with the state vector
$\bPsi=[\psi_{\rm{a}},\psi_{\rme},\psi_{\rm{g}1},\psi_{\rm{g}2}]^T$, and Hamiltonian
\begin{equation}\label{Tr_din_H}
    H = \left(\begin{array}{cccc}
      0 & \Omega_{\rm{p}} & 0 & 0\\
      \Omega_{\rm{p}} & -(\Delta+\rmi\gamma) & \Omega_{\rmd 1} & \Omega_{\rmd 2}\\
      0 & \Omega_{\rmd 1} & 0 & 0 \\
      0 & \Omega_{\rmd 2} & 0 & 0
    \end{array} \right)
\end{equation}
The matrix $A \equiv -\rmi H$ is again the Jacobian of the system.
Here the relation $\lambda=-\rmi\omega$ holds. Solving the
eigenvalues problem (\ref{L_detH}) for the linear tripod, we
obtain two zero eigenvalues,
\begin{equation}\label{om12}
    \omega_{1,2}=0.
\end{equation}
The other two eigenvalues can be found from quadratic equation
\begin{equation}\label{tr_omega_qeq}
    \omega^2+(\Delta+\rmi\gamma)\omega-(\Omega_{\rmd 1}^2+\Omega_{\rmd 2}^2+\Omega_{\rm{p}}^2)=0.
\end{equation}
The eigenvalues $\omega_{3,4}$ must satisfy
\begin{eqnarray}
\omega_3+\omega_4 &=& -(\Delta+\rmi\gamma),\label{Tr_om34a}\\
\omega_3 \omega_4 &=&
-(\Omega_{\rmd 1}^2+\Omega_{\rmd 2}^2+\Omega_{\rm{p}}^2).\label{Tr_om34b}
\end{eqnarray}
We again assume that $\Delta=0$, thus obtaining the following solutions:
\begin{equation}\label{Tr_om34_sol}
    \omega_{3,4} = \{-\rmi\gamma \pm [-\gamma^2 + 4(\Omega_{\rmd 1}^2+\Omega_{\rmd 2}^2+\Omega_{\rm{p}}^2)]^{1/2}\}/2.
\end{equation}
(See also (7) in \cite{unan99}). The Rabi frequencies are chosen in
the form of Gaussian pulses. For $t \rightarrow \pm\infty$ the
Rabi frequencies tend to zero, and from (\ref{Tr_om34a},\ref{Tr_om34b}) it follows
that $\omega_3=0$, $\omega_4 = -\rmi\gamma$. Hence, for $t
\rightarrow \pm\infty$ we have
\begin{equation}\label{Tr_lt_infty}
    \lambda_{1,2,3}=0, \quad \lambda_4=-\gamma.
\end{equation}
However between these two infinite times there is a region where
the Rabi frequencies are large enough, and the discriminant in
(\ref{Tr_om34_sol}) is positive $[D\equiv -\gamma^2 +
4(\Omega_{\rmd 1}^2+\Omega_{\rmd 2}^2+\Omega_{\rm{p}}^2)>0]$. Such a situation
takes place in the interval $t_1<t<t_2$, and from
(\ref{Tr_om34_sol}) we get
\begin{eqnarray}\label{tr_roots34}
\omega_{3,4} &=& -\rmi\gamma/2 \pm \sqrt{D}/2,\\
\lambda_{3,4} &=& -\gamma/2 \mp \rmi\sqrt{D}/2.
\end{eqnarray}
The first two eigenvalues are $\lambda_{1,2}=\omega_{1,2}=0$. The
boundaries $t_1,t_2$ are defined in the same way as in section
\ref{sec:3}.

Hence, in the interval $t_1<t<t_2$, the real parts are $\Re(\lambda_{1,2})=0$,
$\Re(\lambda_{3,4})=-\gamma/2$.

Exactly as in the previous section, we first transform the
variables from the bare states to one bright and two dark states:
\begin{equation}\label{trip_transf}
    (\psi_{\rm{a}}, \psi_{\rme}, \psi_{\rm{g}1}, \psi_{\rm{g}2}) \rightarrow (\psi_{\rm{B}}, \psi_{\rme}, \psi_{\rm{D}1},
    \psi_{\rm{D}2}).
\end{equation}
Parametrizing the Rabi frequencies as
\begin{eqnarray}
    \Omega_{\rmd 2} &=& \Omega\sin{(\phi)},\label{trip_Om_d2}\\
    \Omega_{\rm{p}} &=& \Omega\cos{(\phi)}\sin{(\Theta)}, \label{trip_Om_p}\\
    \Omega_{\rmd 1} &=& \Omega\cos{(\phi)}\cos{(\Theta)},\label{trip_Om_d1}
\end{eqnarray}
we may write down the amplitudes of one bright and two dark
states:
\begin{eqnarray}
    \psi_{\rm{B}} &=& \cos{(\phi)}\sin{(\Theta)}\psi_{\rm{a}}+\cos{(\phi)}\cos{(\Theta)}\psi_{\rm{g}1}+ \nonumber\\
    &+& \sin{(\phi)}\psi_{\rm{g}2},\label{trip_psi_B}\\
    \psi_{\rm{D}1} &=& \cos{(\Theta)}\psi_{\rm{a}}-\sin{(\Theta)}\psi_{\rm{g}1}, \label{trip_psi_D1}\\
    \psi_{\rm{D}2} &=& \sin{(\phi)}\sin{(\Theta)}\psi_{\rm{a}}+\sin{(\phi)}\cos{(\Theta)}\psi_{\rm{g}1}- \nonumber\\
    &-& \cos{(\phi)}\psi_{\rm{g}2}.\label{trip_psi_D2}
\end{eqnarray}
After renormalizing the  time ($t/T \rightarrow t$) and some
rearrangements, we derive the following dynamic system for these
variables:
\begin{equation}\label{trip_lin_varch}
    \rmi\frac{\rmd}{\rmd t}\left(\begin{array}{c}
      \psi_{\rm{B}} \\
      \psi_{\rme} \\
      \psi_{\rm{D}1} \\
      \psi_{\rm{D}2}
    \end{array}\right)=\left(\begin{array}{cccc}
      0 & T\Omega & \alpha_{13} & \alpha_{14} \\
      T\Omega & -T(\Delta+\rmi\gamma) & 0 & 0 \\
      \alpha^{*}_{13} & 0 & 0 & \alpha_{34} \\
      \alpha^{*}_{14} & 0 & \alpha^{*}_{34} & 0
    \end{array}\right)\left(\begin{array}{c}
      \psi_{\rm{B}} \\
      \psi_{\rme} \\
      \psi_{\rm{D}1} \\
      \psi_{\rm{D}2}
    \end{array}\right).
\end{equation}
Here
\begin{eqnarray}
    \alpha_{13} &=& \rmi (\dot{\xi}_{11}\xi_{31}+\dot{\xi}_{13}\xi_{33}),\label{trip_alpha13}\\
    \alpha_{14} &=& \rmi (\dot{\xi}_{11}\xi_{41}+\dot{\xi}_{13}\xi_{43}+\dot{\xi}_{14}\xi_{44}), \label{trip_alpha14}\\
    \alpha_{34} &=&
    \rmi (\dot{\xi}_{31}\xi_{41}+\dot{\xi}_{33}\xi_{43})\label{trip_alpha34},
\end{eqnarray}
and the coefficients $\xi_{ij}$ are defined by the matrix
\begin{equation}\label{trip_lin_ximatr}
    ||\xi_{ij}||=\left(\begin{array}{cccc}
      \cos{(\phi)}\sin{(\Theta)} & 0 & \cos{(\phi)}\cos{(\Theta)} & \sin{(\phi)} \\
      0 & 1 & 0 & 0 \\
      \cos{(\Theta)} & 0 & -\sin{(\Theta)} & 0 \\
      \sin{(\phi)}\sin{(\Theta)} & 0 & \sin{(\phi)}\cos{(\Theta)} &
      -\cos{(\phi)}
    \end{array}\right).
\end{equation}
Note that this matrix realizes the transformation
(\ref{trip_transf}) (see also (\ref{trip_psi_B},\ref{trip_psi_D1},\ref{trip_psi_D2})).

We now reduce the dimension of system (\ref{trip_lin_varch}) in
two steps. For the first step, we eliminate the excited state by
setting
\begin{equation}\label{trip_lin_ad0}
    \dot{\psi}_{\rme} = 0.
\end{equation}
Solving the equation
\begin{equation}\label{trip_ad_psie_eq}
    \Omega\psi_{\rm{B}}-(\Delta+\rmi\gamma)\psi_{\rme}=0,
\end{equation}
(see the second row in (\ref{trip_lin_varch})) for $\psi_{\rme}$, we
get
\begin{equation}\label{trip_lin_ad_psie}
    \psi_{\rme}=\frac{\Omega}{\Delta+\rmi\gamma}\psi_{\rm{B}}.
\end{equation}
Inserting this result in (\ref{trip_lin_varch}), we obtain the
following three-dimensional dynamic system:
\begin{equation}\label{trip_lin_ad}
    \rmi\frac{\rmd}{\rmd t}\left(\begin{array}{c}
      \psi_{\rm{B}} \\
      \psi_{\rm{D}1} \\
      \psi_{\rm{D}2}
    \end{array}\right)      =\left(\begin{array}{ccc}
      \frac{T\Omega^2}{\Delta+\rmi\gamma} & \alpha_{13} & \alpha_{14} \\
      \alpha^{*}_{13} & 0 & \alpha_{34} \\
      \alpha^{*}_{14} & \alpha^{*}_{34} & 0
    \end{array}\right)\left(\begin{array}{c}
      \psi_{\rm{B}} \\
      \psi_{\rm{D}1} \\
      \psi_{\rm{D}2}
    \end{array}\right).
\end{equation}
We can solve this system to find the evolution of $\psi_{\rm{B}}$,
$\psi_{\rm{D}1}$, $\psi_{\rm{D}2}$, and to determine the dynamics of the
excited state using (\ref{trip_lin_ad_psie}).

For the second step, we eliminate the bright state, i.e. we set
\begin{equation}\label{trip_lin_ad20}
    \dot{\psi}_{\rm{B}} = 0,
\end{equation}
in (\ref{trip_lin_ad}). After solving the equation
\begin{equation}\label{trip_ad_psiB_eq}
    \frac{T\Omega^2}{\Delta+\rmi\gamma}\psi_{\rm{B}}+\alpha_{13}\psi_{\rm{D}1}+\alpha_{14}\psi_{\rm{D}2}=0
\end{equation}
for $\psi_{\rm{B}}$ (see the first row in (\ref{trip_lin_ad})), we find
\begin{equation}\label{trip_lin_ad2_psiB}
    \psi_{\rm{B}}=-\frac{\Delta+\rmi\gamma}{T\Omega^2}(\alpha_{13}\psi_{\rm{D}1}+\alpha_{14}\psi_{\rm{D}2}).
\end{equation}
Inserting this result in (\ref{trip_lin_ad}), we obtain the
following second order system:
\begin{equation}\label{trip_lin_ad2}
    \rmi\frac{\rmd}{\rmd t}\left(\begin{array}{c}
      \psi_{\rm{D}1} \\
      \psi_{\rm{D}2}
    \end{array}\right)=H_{2}\left(\begin{array}{c}
      \psi_{\rm{D}1} \\
      \psi_{\rm{D}2}
    \end{array}\right).
\end{equation}
The matrix of this system may be split into two parts:
\begin{equation}\label{trip_lin_A}
    H_{2} = H^{(0)}_{2} + \frac{\Delta+\rmi\gamma}{T\Omega^2}H^{(1)}_{2},
\end{equation}
where
\begin{equation}\label{trip_lin_A0}
    H^{(0)}_{2}=\left(\begin{array}{cc}
      0 & \alpha_{34} \\
      \alpha^{*}_{34} & 0
    \end{array}\right),
\end{equation}
and
\begin{equation}\label{trip_lin_A1}
    H^{(1)}_{2}=-\left(\begin{array}{cc}
      |\alpha_{13}|^2 & \alpha^{*}_{13}\alpha_{14} \\
      \alpha^{*}_{14}\alpha_{13} & |\alpha_{14}|^2
    \end{array}\right).
\end{equation}
When $\Omega^2$ is relatively large, one can neglect the influence
of $H^{(1)}_{2}$ and write approximately $H_{2} \simeq
H^{(0)}_{2}$. Thus one arrives at a simple second order system
\begin{equation}\label{trip_lin_A0_ad2}
    \rmi\frac{\rmd}{\rmd t}\left(\begin{array}{c}
      \psi_{\rm{D}1} \\
      \psi_{\rm{D}2}
    \end{array}\right)=\left(\begin{array}{cc}
      0 & \alpha_{34} \\
      \alpha^{*}_{34} & 0
    \end{array}\right)\left(\begin{array}{c}
      \psi_{\rm{D}1} \\
      \psi_{\rm{D}2}
    \end{array}\right),
\end{equation}
which is equivalent to (27) of \cite{unan98}. However, this system
is not an adiabatically reduced version of the system
(\ref{trip_lin_dina},\ref{trip_lin_dine},\ref{trip_lin_ding1},\ref{trip_lin_ding2}).
Actually, it determines the solution moving exactly on the dark
state manifold comprising the two degenerate states $|D1\rangle$
and $|D2\rangle$ (this statement can be confirmed by applying the
approach of \cite{zhou10}). Comparing (\ref{trip_lin_A}) with the
corresponding result for the linear lambda system,
(\ref{L_din_lin_varch_ad2}), we note that the both Hamiltonians
contain the characteristic time scale $T$ (the pulse width) in the
denominators. In the case of (\ref{trip_lin_A}) this time scale is
involved only with the correction Hamiltonian $H^{(1)}_{2}$, and
it is absent in $H^{(0)}_{2}$ since it corresponds to a zero on
the r.h.s. of (\ref{L_din_lin_varch_ad2}). The correction
$H^{(1)}_{2}$ in (\ref{trip_lin_A}) thus corresponds to the r.h.s.
of (\ref{L_din_lin_varch_ad2}).

\begin{figure}[h!]
\centering\includegraphics[width=0.45\textwidth]{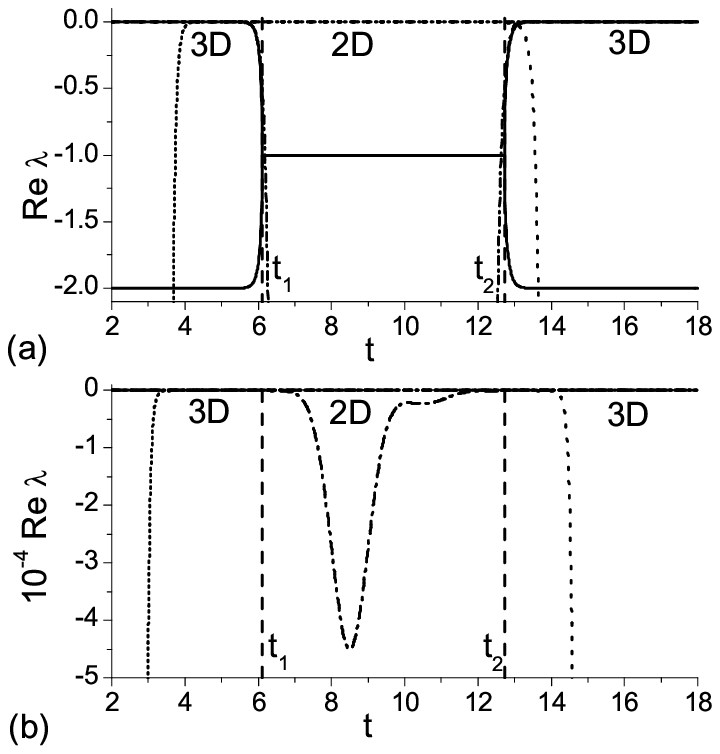}
\caption{\label{Fig_tr_Aties} (a) Dynamics of real parts of
eigenvalues of the Jacobian for linear tripod, computed for all
three cases ($4$D, $3$D and $2$D) by
[(\ref{Tr_om34_sol}),(\ref{om12})], from (\ref{trip_lin_ad}) and
(\ref{trip_lin_A}), respectively. The dashed vertical lines set
the boundaries for the $2$D and $3$D processes. Here $t_1=6.12$
and $t_2=12.73$. The solid lines are the real parts for the $4$D,
the short dash dotted and short dash line correspond to $3$D case,
and the dotted and the short dotted lines are the real parts for
the $2$D reduced system; (b) the dynamics of real parts for the
$3$D and $2$D cases in extended vertical scale. The lines are
chosen in the same way as in (a). The parameters are as follows:
$\Delta = 0$, $\gamma = 2.0$, $\Omega_{0} = 60.0$, $t_{\rm{p}} =
10.7$, $t_{\rmd 1} = 10.0$, $t_{\rmd 2}=8.5$, $K_1=0.75$,
$K_2=5.0$, and $T=1.0$.}
\end{figure}

Figure \ref{Fig_tr_Aties} shows the dynamics of the real parts of
the eigenvalues of the Jacobians for various dimensions. In figure
\ref{Fig_tr_Aties}(a) the solid lines correspond to the exact $4$D
case. For $t<t_{1}$ and $t>t_{2}$ the lower branch, for which $\Re
\lambda \simeq - \gamma$, causes the decay of the excited state.
The upper branch describing the decay rate for the bright state is
small compared to the decay rate of the excited state in these
time intervals indicating that the excited state may be
adiabatically eliminated, whereas the bright state must be left.
In the middle of the passage ($t_{1}<t<t_{2}$) the decay rates of
the excited and bright states become degenerate and equal to
$-\gamma/2$, thus enabling to eliminate them both. The eigenvalues
for the dark states remain both zeros for the whole evolution of
the system demonstrating that the process is adiabatic since the
degenerate dark state does not loose its population.

The dash dotted line in figure \ref{Fig_tr_Aties} displays the
dynamics of the decay rate for the bright state in the $3$D
system. In figure \ref{Fig_tr_Aties}(a) it falls down (grows up)
just after $t_{1}$ (just before $t_{2}$). In figure
\ref{Fig_tr_Aties}(b) the same dynamics is shown in extended
vertical scale. From both figures, \ref{Fig_tr_Aties}(a) and
\ref{Fig_tr_Aties}(b), one may conclude that in the $3$D reduced
system the bright state may be eliminated for $t_{1}<t<t_{2}$, and
it should be preserved for $t<t_{1}$ and $t>t_{2}$, since its
decay rate in the latter case is much less than in the former one.
The two eigenvalues corresponding to degenerate dark states remain
exactly (or almost) zero for the $3$D system. They are small
compared with the decay rate of the bright state. Therefore these
states may not be eliminated for the whole time of evolution.

The dotted and short dotted lines in figure \ref{Fig_tr_Aties}
show the dynamics of the real parts for the $2$D reduced system.
One of them grows rapidly before $t_{1}$ and converges to zero,
whereas the other one is first zero and then decays rapidly after
$t_{2}$. Such a situation suggests that the range of applicability
of the $2$D system should be wider than the time interval
$t_{1}<t<t_{2}$ since the two degenerate dark states survive (do
not decay) when the decay rate is relatively small. However, such
a conclusion would be true if we distributed the whole population
among the degenerate states at the end  of the rapid growth of the
first real part. But if the bright state had some initial
population, it could not be neglected since its real part is close
to zero before $t_{1}$ and after $t_{2}$.

We should also note here that for the $2$D reduced system, the
decay rate for the second dark state grows rapidly from negative
values to zero (the decay rate for the first dark state decreases
rapidly from zero to negative values) in the time intervals
$t<2.0$ ($t>18.0$), i.e. outside of the figure \ref{Fig_tr_Aties}.
But we do not need to take these events into account since we are
interested in the time interval where both decay rates are close
to zero. This interval is determined by the growing (decreasing)
decay rates which are plotted in the figure \ref{Fig_tr_Aties}.

Exactly as in the previous section, we here can also conclude that
the adiabaticity is preserved also for reduced systems, since for
the $3$D ($2$D) approximations the dark states manifold does not
loose its population for the whole time of evolution (in the time
interval that is wider than $t_{1}<t<t_{2}$).

\begin{figure}[h!]
\centering\includegraphics[width=0.45\textwidth]{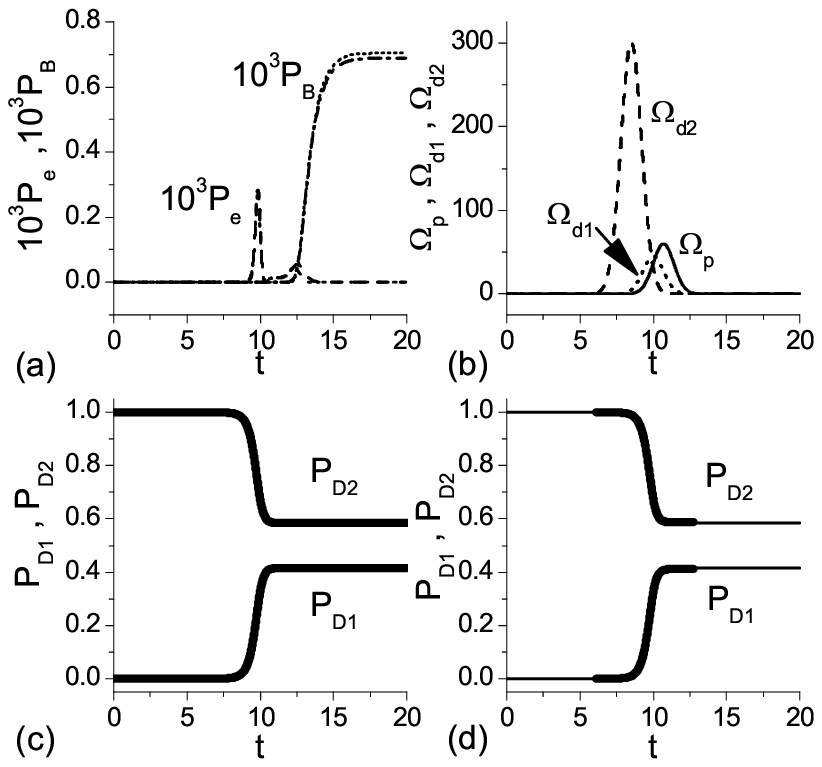}
\caption{\label{Fig_trip_ties} (a) The dynamics of populations
$10^{3}P_{\rme},10^{3}P_{\rm{B}}$ that are computed from
(\ref{trip_lin_dina},\ref{trip_lin_dine},\ref{trip_lin_ding1},\ref{trip_lin_ding2})
(dashed/dash-dotted lines), and from (\ref{trip_lin_ad}),
(\ref{trip_lin_ad_psie}) (dotted/short-dotted lines); (b) the
dynamics of Gaussian pulses computed by
(\ref{trip_lin_Omp},\ref{trip_lin_Omd1},\ref{trip_lin_Omd2});
(c,d) the dynamics of populations of dark states computed in (c)
by
(\ref{trip_lin_dina},\ref{trip_lin_dine},\ref{trip_lin_ding1},\ref{trip_lin_ding2})
(thin solid line), and by (\ref{trip_lin_ad}) (thick solid line);
in (d) the thin solid line represents again the exact solution of
(\ref{trip_lin_dina},\ref{trip_lin_dine},\ref{trip_lin_ding1},\ref{trip_lin_ding2}),
and thick solid line is obtained using the approximation from
(\ref{trip_lin_ad2}) computed in the range of $t_1<t<t_2$. The
parameters are the same as in figure \ref{Fig_tr_Aties}. The
initial conditions are $\psi_{\rm{a}}(0)=1$,
$\psi_{\rm{g}1}(0)=\psi_{\rm{g}2}(0)=\psi_{\rme}(0)=0$.}
\end{figure}

Figure \ref{Fig_trip_ties} displays the results of numerical
computations. In figure \ref{Fig_trip_ties}(a) we have presented
the dynamics of populations of the bright (dashed and dotted
lines), and excited (dashed-dotted and short-dotted lines) states.
The dashed and dotted lines correspond to solutions of exact
equations
(\ref{trip_lin_dina},\ref{trip_lin_dine},\ref{trip_lin_ding1},\ref{trip_lin_ding2}).
The dashed-dotted and short-dotted lines are obtained from
(\ref{trip_lin_ad}) (the bright state), and from
(\ref{trip_lin_ad_psie}) (the excited state). One can see that the
populations of these states remain relatively small
($P_{\rme},P_{\rm{B}} \simeq 10^{-3}$). In figure
\ref{Fig_trip_ties}(b) the dynamics of coupling strengths is
shown. Figures \ref{Fig_trip_ties}(c,d) display the dynamics of
populations of the degenerate dark states. Figure
\ref{Fig_trip_ties}(c) presents the solutions of exact
(\ref{trip_lin_dina},\ref{trip_lin_dine},\ref{trip_lin_ding1},\ref{trip_lin_ding2})
(thin solis line) and that of adiabatically reduced system
(\ref{trip_lin_ad}) (thick solid line). The exact and approximate
solutions are in good quantitative agreement (we do not
distinguish them in the present graph). Both systems were
integrated in the whole time range ($t\in[0.0,20.0]$). In figure
\ref{Fig_trip_ties}(d) we compare the dynamics for exact
(\ref{trip_lin_dina},\ref{trip_lin_dine},\ref{trip_lin_ding1},\ref{trip_lin_ding2})
(thin solid line) with those of adiabatically simplified
(\ref{trip_lin_ad2}) (thick solid line).

\begin{figure}[h!]
\centering\includegraphics[width=0.45\textwidth]{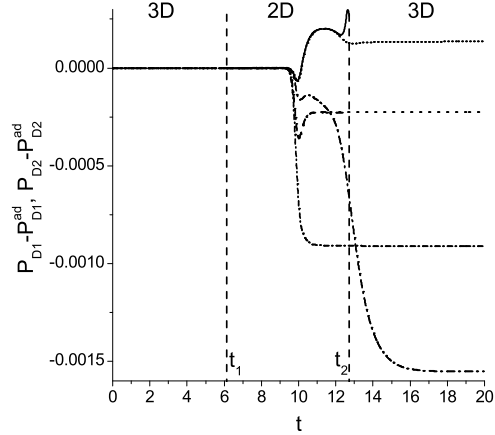}
\caption{\label{Fig_trip_ties_tikr} The differences between the
exact and approximated degenerate dark states; the short dotted
and dotted lines show the differences of
$P_{\rm{D}1}-P^{\rm{ad}}_{\rm{D}1}$ and
$P_{\rm{D}2}-P^{\rm{ad}}_{\rm{D}2}$ for the $3$D system
(\ref{trip_lin_ad}), respectively; the dash dotted and short dash
dotted lines display the corresponding differences for
$P_{\rm{D}1}$ and $P_{\rm{D}2}$ for the $2$D system
(\ref{trip_lin_A0_ad2}), respectively; the solid and dashed lines
show the differences for $P_{\rm{D}1}$ and $P_{\rm{D}2}$ for the
adiabatically reduced system (\ref{trip_lin_ad2}), respectively
(integrated only in the range of $t_1<t<t_2$). The vertical dashed
lines set the boundaries for the $3$D and $2$D processes. The
parameters and initial conditions are the same as in figure
\ref{Fig_trip_ties}.}
\end{figure}

In figure \ref{Fig_trip_ties_tikr} we compare the exactness of
various approximations for the non-linear tripod. We see that the
system (\ref{trip_lin_A0_ad2}) is in a good quantitative agreement
with the exact system, but it is worse than (\ref{trip_lin_ad}) in
the range $t>t_1$. Whereas the $2$D approximation
(\ref{trip_lin_ad2}) coincides with the solution of
(\ref{trip_lin_ad}) in the interval $t_1<t<t_2$ almost
identically. At the end of this interval, just before $t_2$, the
solution of (\ref{trip_lin_ad2}) slightly deviates from that of
(\ref{trip_lin_ad}). This is due to the fact that the magnitude of
element $\frac{\gamma}{T\Omega^2}|\alpha_{13}|^2$ contributed by
the matrix $H^{(1)}_{2}$ in (\ref{trip_lin_A}) becomes large
compared with the magnitude of the element $\alpha_{34}$ of the
matrix $H^{(0)}_{2}$, (\ref{trip_lin_A0}). The remaining elements
of $H^{(1)}_{2}$ are much less than $|\alpha_{13}|^2$ just before
$t_2$.

We should also note that in figure \ref{Fig_trip_ties_tikr} the
solution of "non-exact" $2$D system (\ref{trip_lin_A0}) coincides
almost identically with the exact $4$D solution in the range of
$t<t_1$. However, such a situation takes place when the initial
conditions are very close to the dark state. If we pushed them
away from the dark state, the result of (\ref{trip_lin_A0}) would
become worse than that of the $3$D system (\ref{trip_lin_ad}) (not
shown here). The system thus remains effectively $3$D in the range
of $t<t_1$.

Similarly it can be numerically verified that the solution of
(\ref{trip_lin_ad2}) is not worse than that of (\ref{trip_lin_ad})
just before $t_{2}$ if one pushes the initial conditions further
away from the dark state manifold. Therefore the system remains
$2$D in the whole interval $t_{1}<t<t_{2}$.

\section{\label{sec:4}The non-linear lambda system}

The three-level non-linear Hamiltonian for the non-linear lambda
system (see the figure \ref{Fig_L_scheme}) reads:
\begin{equation}
\label{L_nonl_ham}
\begin{array}{l}
    \hat{H} = -\hbar(\Delta + \rmi\gamma)\hat{\psi}_{\rme}^{\dag}\hat{\psi}_{\rme} +
    \frac{\hbar}{2}(\Omega_{\rm{p}} \hat{\psi}_{\rme}^{\dag}\hat{\psi}_{\rm{a}} \hat{\psi}_{\rm{a}} + \Omega_{\rmd}\hat{\psi}_{\rme}^{\dag}\hat{\psi}_{\rm{g}}\\
    + H.c.).
\end{array}
\end{equation}
Here $\hat{\psi}_{\alpha},\hat{\psi}^{\dag}_{\alpha}$
($\alpha=\rm{a},\rme,\rm{g}$) are the bosonic annihilation and
creation operators for state $\alpha$, respectively. When the
number of particles is much larger than the unity, the boson
operators are replaced by $c$ numbers
$\psi_{\alpha},\psi^{*}_{\alpha}$ (the meanfield treatment
\cite{mackie00,mackie05}) which obey the following Heisenberg
equations
\begin{eqnarray}
    \rmi\dot{\psi}_{\rm{a}} &=& \Omega_{\rm{p}}\psi_{\rm{a}}^{*}\psi_{\rme},\label{L_dina}\\
    \rmi\dot{\psi}_{\rme} &=& -(\Delta + \rmi\gamma)\psi_{\rme} +
    \frac{1}{2}\Omega_{\rm{p}}\psi_{\rm{a}}^{2} +
    \frac{1}{2}\Omega_{\rmd}\psi_{\rm{g}}, \label{L_dine}\\
    \rmi\dot{\psi}_{\rm{g}} &=& \frac{1}{2}\Omega_{\rmd}\psi_{\rme}.\label{L_ding}
\end{eqnarray}
The normalization reads:
\begin{equation}\label{L_nonl_norm}
    |\psi_{\rm{a}}(t)|^{2} + 2[|\psi_{\rm{g}}(t)|^2 + |\psi_{\rme}(t)|^2] \le 1,
\end{equation}
where the equality holds for the initial time.

The nonlinearity enters here in the coupling induced by the pump
field: it couples a particle in the state $e$ with a pair of
particles in the state $a$. Such a nonlinear couplig is
encountered in second harmonic generation in nonlinear optics
(where $a$ represents the fundamental photon and $e$ its second
harmonics), as well as in the photoassociation of atoms into
diatomic molecules \cite{wink05,moal06,mackie00,mackie05}, where
$a$ represents an atomic state, while $e$ and $g$ are excited and
ground diatomic molecular states, respectively.

As for the linear lambda system, we take the Gaussian pulses given
by (\ref{L_lin_Omd},\ref{L_lin_Omp}).

Similarly as in two previous sections, we here define the state vector
$\bPsi=[\psi_{\rm{a}},\psi_{\rme},\psi_{\rm{g}}]^T$. The dynamic system (\ref{L_dina},\ref{L_dine},\ref{L_ding})
can be rewritten in the vector form:
\begin{equation}\label{L_nonl_din_f}
    \rmi\frac{\rmd}{\rmd t}\bPsi=\bi{f}(\bPsi),
\end{equation}
where $\bi{f}$ is the vector of (generally) non-linear functions on
the r.h.s. of system (\ref{L_dina},\ref{L_dine},\ref{L_ding}). The steady state solution $\bPsi_{0}(t)$
of this system represents the dark state which is obtained by solving
\begin{equation}\label{L_nonl_ds}
    \bi{f}(\bPsi_{0})=\mathbf{0}.
\end{equation}
For the non-linear lambda system (\ref{L_dina},\ref{L_dine},\ref{L_ding}) the dark state reads (\cite{mackie00,mackie05}):
\begin{equation}\label{L_nonl_dsol}
    \psi^{0}_{\rm{a}}=\left[\frac{2\Omega_{\rmd}}{\Omega_{\rmd}+\Omega_{\rm{eff}}}\right]^{1/2}, \quad \psi^{0}_{\rme}=0, \quad \psi^{0}_{\rm{g}} = -\frac{2\Omega_{\rm{p}}}{\Omega_{\rmd}+\Omega_{\rm{eff}}},
\end{equation}
with $\Omega_{\rm{eff}}=(\Omega^{2}_{\rmd}+8\Omega^{2}_{\rm{p}})^{1/2}$.

If the solution remains in this state for the whole
time, the adiabatic passage from initially occupied state
$a$ to the target state $g$ takes place provided the
Gaussian pulses $\Omega_{\rmd}(t)$, $\Omega_{\rm{p}}(t)$ arrive in a counter-intuitive
sequence.

We are now interested in the linear stability of the dark state
(\ref{L_nonl_dsol}). To this end, we suppose that the solution of
(\ref{L_nonl_din_f}) evolves in the close neighbourhood of the dark
state $\bPsi_0$, i.e. we express it as a sum
\begin{equation}\label{L_nonl_dev}
    \bPsi(t) = \bPsi_0(t) + \delta\bPsi(t),
\end{equation}
where $\delta\bPsi(t)$ is a deviation of the current solution from
the dark state. Inserting this expression in (\ref{L_nonl_din_f})
yields
\begin{equation}\label{L_nonl_lin}
    \rmi\dot{\bPsi}_{0}(t) + \rmi\frac{\rmd}{\rmd t}\delta\bPsi(t) = \bi{f}(\bPsi_{0}(t)+
    \delta\bPsi(t)).
\end{equation}
Using (\ref{L_nonl_ds}), one finds
\begin{equation}\label{L_nonl_lineq}
    \rmi\frac{\rmd}{\rmd t}\delta\bPsi = M\delta\bPsi - \rmi\dot{\bPsi}_{0}.
\end{equation}
where  $M$ is a matrix with the elements $M_{ij}=\frac{\p f_i}{\p
\psi_j}$ with  $i,j = \rm{a},\rme,\rm{g}$. Here the partial derivatives are
calculated at the dark state. In system (\ref{L_nonl_lineq}) the
non-linear terms have been omitted.

Specifically, for the system (\ref{L_dina},\ref{L_dine},\ref{L_ding}), the matrix $M$
reads
\begin{equation}\label{L_din_M}
    M = \left(\begin{array}{ccc}
      0 & \Omega_{\rm{p}}\psi^{0*}_{\rm{a}} & 0 \\
      \Omega_{\rm{p}}\psi^{0}_{\rm{a}} & -(\Delta+\rmi\gamma) & \Omega_{\rmd}/2 \\
      0 & \Omega_{\rmd}/2 & 0
    \end{array}\right).
\end{equation}
(See also (7) in \cite{pu07}). It is similar to the Hamiltonian
(\ref{L_din_H}) of the linear lambda system. The difference is
that the elements $M_{\rm{ae}}$ and $M_{\rm{ea}}$ contain the component of
the dark state $\psi^{0}_{\rm{a}}$ due to the nonlinearity.

Denoting the Jacobian as $A=-\rmi M$, we can rewrite the
linearised equation (\ref{L_nonl_lineq}) as
\begin{equation}\label{L_nonl_lineqA}
    \frac{\rmd}{\rmd t}\delta\bPsi = A\delta\bPsi - \dot{\bPsi}_{0}.
\end{equation}
In this system, the real parts of the eigenvalues of the Jacobian
$A$ determine the stability of the dark state. If the matrix $M$
has eigenvalues $\omega$, the Jacobian $A$ has the eigenvalues
$\lambda=-\rmi\omega$. In analogy to the linear lambda system, the
eigenvalues $\omega$ can be found from the characteristic equation
\begin{equation}\label{det_M}
    \det||M-I\omega||=0.
\end{equation}
Solving (\ref{det_M}) with (\ref{L_din_M}), we find that one root
is always zero:
\begin{equation}\label{Ln_omega1}
    \omega_1=0.
\end{equation}
The other two eigenvalues obey the quadratic equation:
\begin{equation}\label{Ln_omega_qeq}
    \omega^2+(\Delta+\rmi\gamma)\omega-(\Omega_{\rmd}^2+4\Omega_{\rm{p}}^2|\psi^{0}_{\rm{a}}|^2)/4=0.
\end{equation}
The corresponding eigenvalues $\omega_2$ and  $\omega_3$ obey the following
\begin{eqnarray}
\omega_2+\omega_3 &=& -(\Delta+\rmi\gamma),\label{Ln_om23a}\\
\omega_2 \omega_3 &=&
-(\Omega_{\rmd}^2+4\Omega_{\rm{p}}^2|\psi^{0}_{\rm{a}}|^2)/4.\label{Ln_om23b}
\end{eqnarray}
By setting $\Delta=0$, we get the solutions of quadratic equation:
\begin{equation}\label{Ln_om23_sol}
    \omega_{2,3} = [-\rmi\gamma \pm (-\gamma^2 + \Omega_{\rmd}^2+4\Omega_{\rm{p}}^2|\psi^{0}_{\rm{a}}|^2)^{1/2}]/2.
\end{equation}
(See also the equations under (7) in \cite{pu07}).

\begin{figure}[h!]
\centering\includegraphics[width=0.45\textwidth]{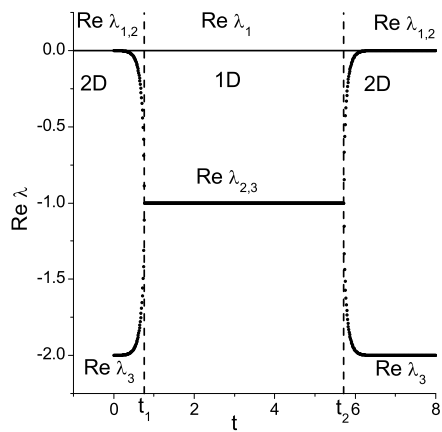}
\caption{\label{Fig_L_Anet}Dynamics of real parts of eigenvalues
of the Jacobian for non-linear lambda system, computed by (\ref{Ln_om23_sol}), and (\ref{Ln_omega1}).
The dashed vertical lines set the boundaries for the $1$D and $2$D
processes. Here $t_1=0.76$ and $t_2=5.71$.  The parameters are as follows: $\Delta = 0$,
$\gamma = 2.0$, $\Omega_{0} = 300.0$, $t_{\rm{p}} = 3.8$, and $t_{\rmd} =
3.0$.}
\end{figure}

In figure \ref{Fig_L_Anet} we show the dynamics of
$\Re(\lambda(t))$ for the non-linear lambda system. One can see
that this picture reproduces the same behaviour as the
corresponding dependence for the linear lambda system shown in
figure \ref{Fig_L_Aties}. This means that the real parts of
eigenvalues of the Jacobian for the nonlinear lambda system are
the same as those for the linear system in the corresponding time
intervals: $t<t_{1}$, $t>t_{2}$ and $t_{1}<t<t_{2}$.

We now adiabatically eliminate the excited state by setting
$\dot{\psi}_{\rme} = 0$. From (\ref{L_dine}) we get
\begin{equation}\label{L_nonl_ad_psie}
    \psi_{\rme} =
    \frac{1}{2(\Delta+\rmi\gamma)}(\Omega_{\rm{p}}\psi^{2}_{\rm{a}}+\Omega_{\rmd}\psi_{\rm{g}}).
\end{equation}
Inserting this result into (\ref{L_dina},\ref{L_ding}) we obtain a second
order system
\begin{eqnarray}
    \rmi\dot{\psi}_{\rm{a}} &=& \Omega_{\rm{p}}\psi_{\rm{a}}^{*}\psi_{\rme},\label{L_nonl_ada}\\
    \rmi\dot{\psi}_{\rm{g}} &=& \frac{1}{2}\Omega_{\rmd}\psi_{\rme}.\label{L_nonl_adg}
\end{eqnarray}
with $\psi_{\rme}$ given by (\ref{L_nonl_ad_psie}).

We now discuss the validity of $3$D and $2$D systems. The $3$D
system (\ref{L_dina},\ref{L_dine},\ref{L_ding}) is valid for all
times. In the ranges $t<t_1$ and $t>t_2$ the $2$D system
(\ref{L_nonl_ada},\ref{L_nonl_adg}) can be applied since there are
two zero real parts $\Re(\lambda_{1,2})=0$, and one negative real
part $\Re(\lambda_3) \simeq -\gamma$. In the range $t_1<t<t_2$ the
process is $1$D since there is only one zero real part
$\Re(\lambda_1)=0$ and two negative real parts
$\Re(\lambda_{2,3})=-\gamma/2$. However, here we do not have any
$1$D equation, one can only propose the $2$D system
(\ref{L_nonl_ada},\ref{L_nonl_adg}). The search for a $1$D system
is a challenging problem.

\begin{figure}[b]
\centering\includegraphics[width=0.45\textwidth]{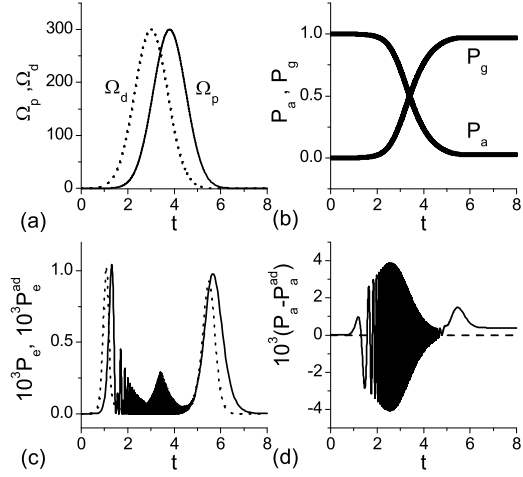}
\caption{\label{Fig_L_net} The dynamics (a) of Gaussian pulses
(\ref{L_lin_Omd},\ref{L_lin_Omp}), (b) of populations
$P_{\rm{a}}$, $P_{\rm{g}}$, (c) of population $P_{\rme}$ in
enlarged scale, and (d) the difference of $P_{\rm{a}}$ between
exactly ($P_{\rm{a}}$) and adiabatically ($P^{\rm{ad}}_{\rm{a}}$)
obtained solutions. In (b,c) the solid thin lines represent the
solutions of exact (\ref{L_dina},\ref{L_dine},\ref{L_ding}), and
the (solid thick/dotted) lines are the solutions of adiabatically
reduced system (\ref{L_nonl_ada},\ref{L_nonl_adg}) with excited
state given by (\ref{L_nonl_ad_psie}). The parameters are the same
as in figure \ref{Fig_L_Anet}. The initial conditions are
$\psi_{\rm{a}}(0)=1$, $\psi_{\rm{g}}(0)=\psi_{\rme}(0)=0$.}
\end{figure}

In the figure \ref{Fig_L_net} we have plotted the relevant
dynamics for the case of adiabatically reduced non-linear lambda
system. In the figure \ref{Fig_L_net}(a) we show the Gaussian
pulses that are ordered counter-intuitively. From figure
\ref{Fig_L_net}(b) one may conclude that the solutions of
adiabatically reduced system are in good quantitative agreement
with the solutions of exact system (we do not distinguish them in
the figure). In figure \ref{Fig_L_net}(c) we see that the
difference for $P_{\rme}$ between exact and approximated solutions
is significant. It can be explained by the fact that the magnitude
of probability $P_{\rme}$ is small. On the other hand, the
difference in the case of $P_{\rm{a}}$, $P_{\rm{g}}$ is of the
same order but we do not distinguish it, since the magnitudes of
these quantities are much larger. In figure \ref{Fig_L_net}(d) we
have plotted the dynamics of the difference between the exact
($P_{\rm{a}}$) and adiabatically reduced ($P^{\rm{ad}}_{\rm{a}}$)
solutions of the population $P_{\rm{a}}$. (See also the blue line
in figure 2(c) in \cite{itin07}). The difference is of the same
order as that in figure \ref{Fig_L_net}(c).

\begin{figure}[b]
\centering\includegraphics[width=0.45\textwidth]{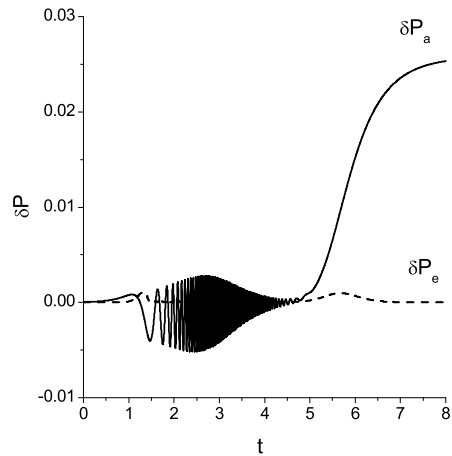}
\caption{\label{Fig_L_net_plg} The dynamics of the differences of
populations between the current state and the dark state, for
$\delta P_{\rm{a}}(t) =
|\psi_{\rm{a}}(t)|^{2}-|\psi^{0}_{\rm{a}}(t)|^{2}$ (solid line),
and $\delta P_{\rme}(t) = 2|\psi_{\rme}(t)|^{2}$ (dashed line) for
the nonlinear lambda system. The dynamics of
$\psi_{\rm{a}}(t),\psi_{\rme}(t)$ are computed from
(\ref{L_dina},\ref{L_dine},\ref{L_ding}), and
$\psi^{0}_{\rm{a}}(t)$ is given by (\ref{L_nonl_dsol}).  The
parameters are the same as in figure \ref{Fig_L_Anet}. The initial
conditions are $\psi_{\rm{a}}(0)=1$,
$\psi_{\rm{g}}(0)=\psi_{\rme}(0)=0$.}
\end{figure}

In figure \ref{Fig_L_net_plg} we have plotted the dynamics of
differences between the populations of the current state and
corresponding dark state. The difference for the initial ground
state $a$ deviates up to $0.03$ at the end of the passage. Whereas
the difference for the excited state remains much less. The
corresponding difference for state $g$ is almost symmetric to that
of the state $a$ in respect to the zero (not shown here). We
conclude that the process is adiabatic since the solution remains
in a close neighborhood to the dark state.

\section{\label{sec:5}The non-linear tripod}

We consider the atom-molecule transition in ultracold quantum
gases via photoassociation. This problem has potential
applications during the creation of ultracold molecules and
quantum superchemistry. The underlying physics is closely related
to the STIRAP and has been widely studied in the context of atomic
physics and quantum optics \cite{hioe84,vitan01a}.

The level structure of the atom-molecule tripod system is shown in
figure \ref{Fig_tr_scheme}. In ultracold atomic systems, the level
$a$ denotes the atomic Bose-Einstein condensation (BEC), which
couples an excited state of a diatomic molecular BEC via the pump
field $\Omega_{\rm{p}}(t)$. Such an excited state is represented
by high-lying vibration levels of the single excited molecule. The
excited state is coupled to the two ground states of the molecular
BEC, $g1$ and $g2$, with the strengths $\Omega_{\rmd 1}(t)$ and
$\Omega_{\rmd 2}(t)$, respectively.

Assuming a two-photon resonance condition, the four-level
non-linear Hamiltonian for non-linear tripod takes the form:
\begin{equation}
\label{trip_ham}
\begin{array}{l}
    \hat{H} = -\hbar(\Delta + \rmi\gamma)\hat{\psi}_{\rme}^{\dag}\hat{\psi}_{\rme} +
    \frac{\hbar}{2}(\Omega_{\rm{p}} \hat{\psi}_{\rme}^{\dag}\hat{\psi}_{\rm{a}} \hat{\psi}_{\rm{a}} + \Omega_{\rmd 1}\hat{\psi}_{\rme}^{\dag}\hat{\psi}_{\rm{g}1}
    +  \\ \Omega_{\rmd 2}\hat{\psi}_{\rme}^{\dag}\hat{\psi}_{\rm{g}2} + H.c.).
\end{array}
\end{equation}
Here $\hat{\psi}_{\alpha},\hat{\psi}^{\dag}_{\alpha}$
($\alpha=\rm{a},\rme,\rm{g}1,\rm{g}2$) are the bosonic
annihilation and creation operators for state $\alpha$,
respectively. To explore the behaviour of the system under time
evolution, we consider the problem under mean-field approximation,
which is reasonable for bosonic systems when the number of
particles is large compared with unity \cite{mackie00,mackie05}.
In this limit, the bosonic operators are replaced by $c$ numbers,
and the Heisenberg equation leads to the following equations of
motion for the probability amplitudes:
\begin{eqnarray}
    \rmi\dot{\psi}_{\rm{a}} &=& \Omega_{\rm{p}}\psi_{\rm{a}}^{*}\psi_{\rme},\label{trip_dina}\\
    \rmi\dot{\psi}_{\rme} &=& -(\Delta + \rmi\gamma)\psi_{\rme} +
    \frac{1}{2}\Omega_{\rm{p}}\psi_{\rm{a}}^{2} +
    \frac{1}{2}\Omega_{\rmd 1}\psi_{\rm{g}1} +\nonumber\\
    &+& \frac{1}{2}\Omega_{\rmd 2}\psi_{\rm{g}2}, \label{trip_dine}\\
    \rmi\dot{\psi}_{\rm{g}1} &=& \frac{1}{2}\Omega_{\rmd 1}\psi_{\rme},\label{trip_ding1}\\
    \rmi\dot{\psi}_{\rm{g}2} &=& \frac{1}{2}\Omega_{\rmd 2}\psi_{\rme}.\label{trip_ding2}
\end{eqnarray}
The nonlinear term enters here when the molecules are obtained via
associating cold atoms.

The normalization reads:
\begin{equation}\label{trip_norm}
    |\psi_{\rm{a}}(t)|^{2} + 2[|\psi_{\rm{g}1}(t)|^2 + |\psi_{\rm{g}2}(t)|^2 +
    |\psi_{\rme}(t)|^2] \le 1,
\end{equation}
where the equality holds for the initial time.

Like in the case of linear tripod, the Gaussian pulses are given
by (\ref{trip_lin_Omp},\ref{trip_lin_Omd1},\ref{trip_lin_Omd2}).

\begin{figure}[h!]
\centering\includegraphics[width=0.45\textwidth]{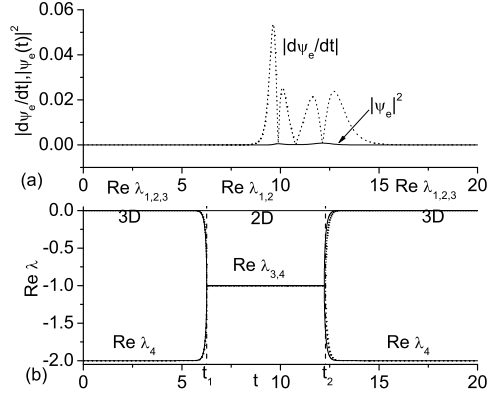}
\caption{\label{Fig_tr_non_eigvs} (a) Dynamics of
$|\dot{\psi}_{\rme}(t)|$ (dotted), and $|\psi_{\rme}(t)|^{2}$
(solid) for non-linear tripod, both computed by
(\ref{trip_dina},\ref{trip_dine},\ref{trip_ding1},\ref{trip_ding2});
(b) dynamics of real parts of eigenvalues of the Jacobian for
non-linear tripod, computed by (\ref{Tr_nonl_om34_sol}), and
(\ref{tr_nonl_om12}). The solid lines are the exact roots, and the
dotted lines correspond to stability of the current solution (see
in the text for details). The dashed black vertical lines set the
boundaries for the $3$D and $2$D processes. Here $t_1=6.26$ and
$t_2=12.29$. The parameters are as follows: $\Delta = 0$, $\gamma
= 2.0$, $\Omega_{0} = 60.0$, $t_{\rm{p}} = 10.7$, $t_{\rmd 1} =
10.0$, $t_{\rmd 2}=8.5$, $K_1=0.75$, and $K_2=5.0$. The initial
conditions are $\psi_{\rm{a}}(0)=1$,
$\psi_{\rm{g}1}(0)=\psi_{\rm{g}2}(0)=\psi_{\rme}(0)=0$.}
\end{figure}

Similar to the linear case, we define the following state vector
of the system: $\bPsi=[\psi_{\rm{a}},\psi_{\rme},\psi_{\rm{g}1},\psi_{\rm{g}2}]^T$.
The dynamic equations (\ref{trip_dina},\ref{trip_dine},\ref{trip_ding1},\ref{trip_ding2}) can be rewritten in a
vector form given by (\ref{L_nonl_din_f}), where $\bi{f}$ is now
the vector of non-linear functions on the r.h.s of
(\ref{trip_dina},\ref{trip_dine},\ref{trip_ding1},\ref{trip_ding2}). The manifold of the steady states of this system
represents the dark state
$\bPsi_0=[\psi^{0}_{\rm{a}},\psi^{0}_{\rme},\psi^{0}_{\rm{g}1},\psi^{0}_{\rm{g}2}]^T$.
This manifold has to satisfy (\ref{L_nonl_ds}) and the condition
of normalization. Since $\psi^{0}_{\rme}=0$, the dark state obeys the
following:
\begin{eqnarray}
   \Omega_{\rm{p}}(\psi^{0}_{\rm{a}})^2 + \Omega_{\rmd 1}\psi^{0}_{\rm{g}1} + \Omega_{\rmd 2}\psi^{0}_{\rm{g}2} = 0,\label{trip_dsa}\\
   |\psi^{0}_{\rm{a}}|^{2} + 2|\psi^{0}_{\rm{g}1}|^{2} + 2|\psi^{0}_{\rm{g}2}|^{2} = 1,\label{trip_dsb}\\
    \psi^{0}_{\rme}  = 0.\label{trip_dsc}
\end{eqnarray}
We are interested in the linear stability of this dark state.
Therefore we suppose that the solution of
(\ref{trip_dina},\ref{trip_dine},\ref{trip_ding1},\ref{trip_ding2})
evolves in the close neighbourhood of the dark state $\bPsi_0$,
i.e. we express it as a sum given by (\ref{L_nonl_dev}) where
$\delta\bPsi(t)$ is the deviation of the current solution from the
dark state. Hence one arrives at a linearised equation similar to
the one for the non-linear lambda system (\ref{L_nonl_lineq})
where the matrix $M$ reads:
\begin{equation}\label{trip_nonl_M}
    M = \left(
          \begin{array}{cccc}
            0 & \Omega_{\rm{p}}\psi^{0*}_{\rm{a}} & 0 & 0 \\
            \Omega_{\rm{p}}\psi^{0}_{\rm{a}} & -[\Delta+\rmi\gamma] & \Omega_{\rmd 1}/2 & \Omega_{\rmd 2}/2 \\
            0 & \Omega_{\rmd 1}/2 & 0 & 0 \\
            0 & \Omega_{\rmd 2}/2 & 0 & 0 \\
          \end{array}
        \right).
\end{equation}
Note that this matrix is very similar to the Hamiltonian
(\ref{Tr_din_H}) for the linear tripod. The main difference
between them is dependence of $M$ on $\psi^{0}_{\rm{a}}$, that arises due
to the nonlinearity. On the other hand, this matrix is also
similar with corresponding matrix for the non-linear lambda system.
The Jacobian of the linearised system is given by $A=-\rmi M$. The
eigenvalues $\omega$ of the matrix $M$ correspond to eigenvalues
$\lambda=-\rmi\omega$ of Jacobian. Solving the eigenvalues problem
for matrix $M$, we get two zero eigenvalues:
\begin{equation}\label{tr_nonl_om12}
    \omega_{1,2}=0.
\end{equation}
The other two eigenvalues can be found from the quadratic equation
\begin{equation}\label{tr_nonl_omega_qeq}
    \omega^2+(\Delta+\rmi\gamma)\omega-(\Omega_{\rmd 1}^2+\Omega_{\rmd 2}^2+4\Omega_{\rm{p}}^{2}|\psi^{0}_{\rm{a}}|^{2})/4=0.
\end{equation}
The eigenvalues $\omega_{3,4}$ satisfy the condition
\begin{eqnarray}
\omega_3+\omega_4 &=& -(\Delta+\rmi\gamma),\label{Tr_nonl_om34a}\\
\omega_3 \omega_4 &=&
-(\Omega_{\rmd 1}^2+\Omega_{\rmd 2}^2+4\Omega_{\rm{p}}^{2}|\psi^{0}_{\rm{a}}|^{2})/4.\label{Tr_nonl_om34b}
\end{eqnarray}
We again assume that $\Delta=0$, thus obtaining the following solutions:
\begin{equation}\label{Tr_nonl_om34_sol}
    \omega_{3,4} = [-\rmi\gamma \pm (-\gamma^2 + \Omega_{\rmd 1}^2+\Omega_{\rmd 2}^2+4\Omega_{\rm{p}}^{2}|\psi^{0}_{\rm{a}}|^{2})^{1/2}]/2.
\end{equation}
In figure \ref{Fig_tr_non_eigvs}(b) we plot the dynamics of
eigenvalues of Jacobian for the non-linear tripod. The way of
finding the eigenvalues is discussed below. As we saw above, the
behaviours of corresponding eigenvalues for the linear and
non-linear lambda systems was the same. Comparing the figures
\ref{Fig_tr_non_eigvs}(b) and \ref{Fig_tr_Aties} we see that here
one can make an identical conclusion: the roots behave in the same
manner for the linear and non-linear tripods.

Now we discuss the computing the dynamics of the real parts of
eigenvalues, i.e. $\Re(\lambda) = \Re(\lambda(t))$. The matrix $M$
in the present case depends on $\psi^{0}_{\rm{a}}$, (see
$M_{\rm{ae}}$ and $M_{\rm{ea}}$ in (\ref{trip_nonl_M})). In the
case of non-linear lambda system, the dark state was uniquely
defined as a function of Rabi frequencies, (\ref{L_nonl_dsol}).
However, in the present case, for non-linear tripod, the dark
state is a manifold that is given by
(\ref{trip_dsa},\ref{trip_dsb},\ref{trip_dsc}). But we need a
definite function of time $\psi^{0}_{\rm{a}}=\psi^{0}_{\rm{a}}(t)$
in order to get the dynamics of eigenvalues. Therefore we use the
parametrization of the dark state that was derived in
\cite{zhou10}. If the solution of
(\ref{trip_dina},\ref{trip_dine},\ref{trip_ding1},\ref{trip_ding2})
evolves on the dark state manifold, we may express that solution
in terms of only two variables (parameters), $[u_{1}(t),
u_{2}(t)]$:
\begin{eqnarray}
\psi_{\rm{a}}^{0} &=& \left[\frac{\delta_{\rm{p}}}{\cos{(\Theta)}}\right]^{1/2}u_{2}^{1/2},\label{a0}\\
\psi_{\rm{g}1}^{0} &=& u_{1}\sin{(\Theta)} - u_{2}\cos{(\Theta)},\label{g1}\\
\psi_{\rm{g}2}^{0} &=& -u_{1}\cos{(\Theta)} - u_{2}\sin{(\Theta)}.\label{g2}
\end{eqnarray}
(See the system of equations before (17) in \cite{zhou10}). Here
$\delta_{\rm{p}}=\Omega_{\rmd 1}/\Omega_{\rm{p}}$ and $\Theta$ is defined by
$\tan{(\Theta)}=\Omega_{\rmd 2}/\Omega_{\rmd 1}$. In \cite{zhou10} it was
also shown that in the adiabatic limit, the parameters should obey
the equations (see (17) in \cite{zhou10})

\begin{eqnarray}
  \dot{u}_{1}+\dot{\Theta}u_{2} = 0,\hspace{30mm} \label{u1}\\
\dot{u}_{2}\left(1+\frac{\delta_{\rm{p}}}{4u_{2}\cos{(\Theta)}}\right)-\dot{\Theta}u_{1}+
\frac{\rmd}{\rmd
t}\left[\frac{\delta_{\rm{p}}}{4\cos{(\Theta)}}\right] = 0.\label{u2}
\end{eqnarray}

We integrate the system (\ref{u1},\ref{u2}) and insert its
solution in the parametrization (\ref{a0}) thus obtaining
the necessary dynamics of $\psi_{\rm{a}}^{0}(t)$. After inserting this
dynamics in (\ref{Tr_nonl_om34_sol}) we get the dynamics of
eigenvalues of the Jacobian for the non-linear tripod system.

We may also suppose that the deviation for the amplitude $\psi_{\rm{a}}$
is almost zero, $\delta\psi_{\rm{a}}(t) \simeq 0$. We thus can make a
substitution in (\ref{Tr_nonl_om34a},\ref{Tr_nonl_om34b}) and (\ref{Tr_nonl_om34_sol}):
\begin{equation}\label{trip_nonl_subs}
    \psi^{0}_{\rm{a}} \rightarrow \psi_{\rm{a}}.
\end{equation}
Subsequently one can numerically solve the system
(\ref{trip_dina},\ref{trip_dine},\ref{trip_ding1},\ref{trip_ding2}). By inserting $\psi_{\rm{a}}(t)$ in (\ref{Tr_nonl_om34a},\ref{Tr_nonl_om34b})
and (\ref{Tr_nonl_om34_sol}), one gets the approximate dynamics of
the eigenvalues. Actually, this approach means the analysis of the
stability of the current solution
$\bPsi(t)=[\psi_{\rm{a}}(t),\psi_{\rme}(t),\psi_{\rm{g}1}(t),\psi_{\rm{g}2}(t)]^T$.
The dynamics of $|\dot{\psi}_{\rme}(t)|$ and $|\psi_{\rme}(t)|^{2}$ is
plotted in figure \ref{Fig_tr_non_eigvs}(a). The first dynamics
indicates that the magnitude of the r.h.s. of (\ref{trip_dine}) is
of the order of $0.06$. The second dynamics shows that the excited
level remains almost unpopulated throughout the passage. We
therefore conclude that the process is almost adiabatic, and one
may justify the substitution (\ref{trip_nonl_subs}).

In figure \ref{Fig_tr_non_eigvs}(b) we plot the dynamics of
$\Re(\lambda(t))$ computed by the both ways. The solid line shows
the dynamics of $\Re(\lambda(t))$ computed by using the exact
value of $\psi^{0}_{\rm{a}}(t)$, and the dotted line displays the
approximate dynamics that is obtained by using the substitution
(\ref{trip_nonl_subs}). We can see from figure
\ref{Fig_tr_non_eigvs}(b) that the stability of the current
solution (dotted line) is identical to that of the dark state at
the beginning and the middle of the process. However, the
splitting of the real parts for the approximate eigenvalues is
slightly delayed with respect to the exact ones. The good
quantitative agreement of the both results confirms the validity
of the approximation (\ref{trip_nonl_subs}); it also shows that
the current solution evolves in the close neighbourhood of the
moving dark state (\ref{a0},\ref{g1},\ref{g2}).

Exactly as in the previous sections, we adiabatically eliminate
the excited state by setting $\dot{\psi}_{\rme} = 0$. From
(\ref{trip_dine}) we get
\begin{equation}\label{trip_nonl_ad_psie}
    \psi_{\rme} =
    \frac{1}{2(\Delta+\rmi\gamma)}(\Omega_{\rm{p}}\psi^{2}_{\rm{a}}+\Omega_{\rmd 1}\psi_{\rm{g}1}+\Omega_{\rmd 2}\psi_{\rm{g}2}).
\end{equation}
Inserting this expression in (\ref{trip_dina},\ref{trip_ding1},\ref{trip_ding2}), we obtain
\begin{eqnarray}
    \rmi\dot{\psi}_{\rm{a}} &=& \Omega_{\rm{p}}\psi_{\rm{a}}^{*}\psi_{\rme},\label{trip_nonl_ada}\\
    \rmi\dot{\psi}_{\rm{g}1} &=& \frac{1}{2}\Omega_{\rmd 1}\psi_{\rme},\label{trip_nonl_adg1} \\
    \rmi\dot{\psi}_{\rm{g}2} &=& \frac{1}{2}\Omega_{\rmd 2}\psi_{\rme}.\label{trip_nonl_adg2}
\end{eqnarray}
In these equations we use the expression of $\psi_{\rme}$ given by
(\ref{trip_nonl_ad_psie}).

\begin{figure}[b]
\centering\includegraphics[width=0.45\textwidth]{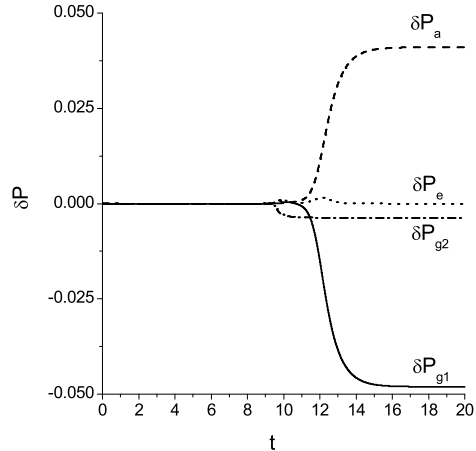}
\caption{\label{Fig_tr_net_plg} The dynamics of the differences of
populations between the current state and the dark state for the
nonlinear tripod, $\delta P_{a}(t) =
|\psi_{\rm{a}}(t)|^{2}-|\psi^{0}_{\rm{a}}(t)|^{2}$ (dashed line),
$\delta P_{\rme}(t) = 2|\psi_{\rme}(t)|^{2}$ (dotted line),
$\delta P_{\rm{g}1}(t) =
2|\psi_{\rm{g}1}(t)|^{2}-2|\psi^{0}_{\rm{g}1}(t)|^{2}$ (solid
line), and $\delta P_{\rm{g}2}(t) =
2|\psi_{\rm{g}2}(t)|^{2}-2|\psi^{0}_{\rm{g}2}(t)|^{2}$ (dash
dotted line). The dynamics of
$\psi_{\rm{a}}(t),\psi_{\rme}(t),\psi_{\rm{g}1}(t)$ and
$\psi_{\rm{g}2}(t)$ are computed from
(\ref{trip_dina},\ref{trip_dine},\ref{trip_ding1},\ref{trip_ding2}),
and
$\psi^{0}_{\rm{a}}(t),\psi^{0}_{\rm{g}1}(t),\psi^{0}_{\rm{g}2}(t)$
are found from (\ref{a0},\ref{g1},\ref{g2}) and
(\ref{u1},\ref{u2}). The parameters are the same as in figure
\ref{Fig_tr_non_eigvs}. The initial conditions are
$\psi_{\rm{a}}(0)=1$,
$\psi_{\rm{g}1}(0)=\psi_{\rm{g}2}(0)=\psi_{\rme}(0)=0$.}
\end{figure}

In figure \ref{Fig_tr_net_plg} we have depicted the dynamics of
the deviations of the current populations from those of the dark
state. One can see that all the three populations deviate up to
$0.05$ showing that the solution remains in a proximity to the
dark state manifold. Therefore one may conclude that the process
is adiabatic.

\begin{figure}[h!]
\centering\includegraphics[width=0.45\textwidth]{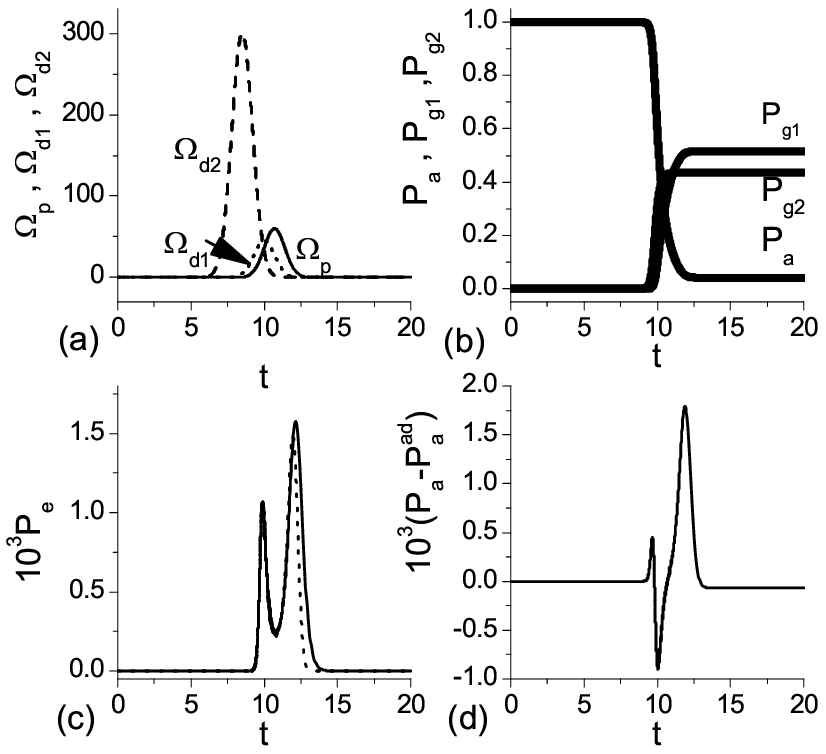}
\caption{\label{Fig_trip_net} The dynamics (a) of Gaussian pulses
(\ref{trip_lin_Omp},\ref{trip_lin_Omd1},\ref{trip_lin_Omd2}), (b)
of populations $P_{\rm{a}}$, $P_{\rm{g}1}$, $P_{\rm{g}2}$, and (c)
of population $P_{\rme}$ in enlarged scale. In (d) it is shown the
difference between the exact ($P_{\rm{a}}$ from
(\ref{trip_dina},\ref{trip_dine},\ref{trip_ding1},\ref{trip_ding2}))
and adiabatically reduced ($P^{\rm{ad}}_{\rm{a}}$ from
(\ref{trip_nonl_ada},\ref{trip_nonl_adg1},\ref{trip_nonl_adg2}))
populations in enlarged scale. In (b,c) the solid thin lines
represent the solutions of exact
(\ref{trip_dina},\ref{trip_dine},\ref{trip_ding1},\ref{trip_ding2}),
and the (solid thick/dotted) lines are the solutions of
adiabatically reduced system
(\ref{trip_nonl_ada},\ref{trip_nonl_adg1},\ref{trip_nonl_adg2})
with excited state given by (\ref{trip_nonl_ad_psie}). The
parameters are as follows: $\Delta = 0$, $\gamma = 2.0$,
$\Omega_{0} = 60.0$, $t_{\rm{p}} = 11.5$, $t_{\rmd 1} = 10.0$,
$t_{\rmd 2}=8.5$, $K_1=0.75$, and $K_2=5.0$. The initial
conditions are the same as in figure \ref{Fig_tr_non_eigvs}.}
\end{figure}

In figure \ref{Fig_trip_net} we have plotted the dynamics for the
case of non-linear tripod. Figure \ref{Fig_trip_net}(a) shows a
sequence of Gaussian pulses. In figure \ref{Fig_trip_net}(b) we
show the dynamics of populations of the levels. The solutions of
approximated system are in good quantitative agreement with those
of the exact system (in the figure we do not distinguish them).
From figure \ref{Fig_trip_net}(c) we see that the population of
the excited state is reproduced by the approximated system with a
significant error. Again, as in the previous section, we explain
this fact by small magnitude of quantity $P_{\rme}$. In figure
\ref{Fig_trip_net}(d) we have plotted the difference between exact
($P_{\rm{a}}$) and adiabatically approximated
($P^{\rm{ad}}_{\rm{a}}$) population $P_{\rm{a}}$. It is of the
same order as in the case of excited state (see figure
\ref{Fig_trip_net}(c)).

\section{\label{sec:6} Some remarks about the one-photon detuning}

Up to now we have been setting $\Delta=0$ for all considered
systems. In this section we shall explore a behaviour of the system
in the presence of the non-zero one-photon detuning $\Delta \neq
0$. In that case, when solving the quadratic equations for the
eigenvalues of the Jacobians, one gets the complex-valued
discriminants, $D \equiv |D|\rme^{\rmi\varphi}$, where $|D|$ is their
real amplitude, and $\varphi$ is their phase. The eigenvalues of
the Jacobian then read:
\begin{eqnarray}
     \lambda_{3,4} = \left[-\frac{1}{2}\gamma \pm \frac{1}{2}|D|^{1/2}\sin{\left(\frac{\varphi}{2}\right)}\right] + \nonumber \\ + \rmi\left[\frac{1}{2}\Delta \mp \frac{1}{2}|D|^{1/2}\cos{\left(\frac{\varphi}{2}\right)}\right].
\end{eqnarray}
Here the real and imaginary parts of the discriminant are given by
\begin{eqnarray}
    \Re (D) &=& \Delta^{2}-\gamma^{2}+\Omega^{2}_{\rmd 1}+\Omega^{2}_{\rmd 2}+4\Omega^{2}_{\rm{p}}|\psi^{0}_{\rm{a}}|^{2},\label{trip_nonl_Dre}\\
    \Im (D) &=& 2\gamma\Delta.\label{trip_nonl_Dim}
\end{eqnarray}
These equations are valid for the non-linear tripod; for the other
systems we get the similar expressions. If $\Delta=0$ (as assumed
previously), the imaginary part becomes zero, i.e. $\Im (D)=0$;
the phase may be either $0$ or $\pi$; in the interval
$t_{1}<t<t_{2}$, it is $\varphi=0$, and in the ranges $t<t_{1}$,
$t>t_{2}$ it is $\varphi=\pi$. For $\varphi=0$ we have $\Re
(\lambda_{3,4})=-\gamma/2$, and for $\varphi=\pi$ we get $\Re
(\lambda_{3})\simeq 0$, $\Re (\lambda_{4}) \simeq -\gamma$. We
have got these results for all the considered systems (see e.g.
figure \ref{Fig_tr_non_eigvs}(b)). However, in the case of the
non-zero one-photon detunings, in the interval $t_{1}<t<t_{2}$,
these real parts are no longer coinciding; they are symmetrically
surrounding the value $-\gamma/2$, and the difference between them
becomes equal to
\begin{eqnarray}
    \Re \lambda_{3} - \Re \lambda_{4} = |D|^{1/2}\sin{\left(\frac{\varphi}{2}\right)} \simeq  \nonumber\\ \simeq \frac{\Im (D)}{2[\Re (D)]^{1/2}}.
\end{eqnarray}
The latter approximation is valid for the small values of $\Im (D)
<< \Re (D)$. Such a situation takes place in the middle of the
passage, when the Rabi frequencies are large compared to the
one-photon detuning and losses.

We thus conclude that for such small detunings the difference
between the negative real parts remains small, and our statements
about the reduction of dimension remain valid.


\section{\label{sec:7}Conclusions}

We have analysed the adiabatic reduction of the
dimension of the linear and non-linear three- and four-level
systems. By evaluating the corresponding Jacobians and computing
the dynamics of real parts of their eigenvalues (the non-zero
eigenvalues are found from quadratic characteristic equations),
one may define the dimensionality of the processes. This
dimensionality is given by the number of zero real parts since the
negative real parts cause the contraction of the nearby solutions
towards the dark state. At the beginning and the end of the
dynamics, there is always only one negative real part. Hence one
may eliminate only one state representing the excited state. In
the middle of the process, one of the zero real parts becomes
negative thus making the number of negative real parts equal to
two. In this time interval we may eliminate two variables
corresponding to the excited and bright states respectively. For
linear systems, we have eliminated both excited and bright states.
However, for non-linear systems, we have restricted ourselves by
eliminating the excited state. This is due to the fact that the
definition of a bright state for the non-linear systems is not
available. We suppose that the remaining stable degrees of
freedom in the non-linear systems can be eliminated by using the
asymptotic methods of non-linear dynamics.

The main finding of this work is revealing that the whole STIRAP
evolution for all considered systems is divided into three time
intervals with different number of the negative real parts of the
Jacobians. The evolution of the real parts is equivalent for the
corresponding linear and non-linear systems (as one can see in
figures \ref{Fig_L_Aties},\ref{Fig_tr_Aties},\ref{Fig_L_Anet}, and
\ref{Fig_tr_non_eigvs}(b)). This suggests that the non-linear
systems may be potentially reduced as the linear ones. Physically
this means that the considered three/four-level schemes may be
regarded as schemes with lower dimension, i.e. with fewer levels
involved. In the time intervals $t<t_{1}$ and $t>t_{2}$, the
initially three-level system is effectively a two-level one, and
in the range of $t_{1}<t<t_{2}$ it contains a single level.
Analogously, the initially four-level scheme may be regarded as a
three-level (two-level) system in the time ranges, where
$t<t_{1}$, $t>t_{2}$ ($t_{1}<t<t_{2}$).

A sensitive problem is the definition of a dark state for the
non-linear tripod. In the case of non-linear lambda system, the
dark state is a moving point (\ref{L_nonl_dsol}) in the phase
space. However, for the non-linear tripod we have a manifold
(\ref{trip_dsa},\ref{trip_dsb},\ref{trip_dsc}) of dark states. If
one wishes to get the dynamics of real parts of eigenvalues of the
Jacobian, one needs a definite value of complex amplitude
$\psi^{0}_{\rm{a}}$ belonging to the manifold. We here use two
ways for the stability analysis of the dark state. The first way
is to parametrize the dark state manifold by using the method
developed in \cite{zhou10}. This method enables one to find the
definite dynamics of $\psi^{0}_{\rm{a}}(t)$. We thus managed to
find the exact dynamics of eigenvalues. The second way is to
simply substitute using (\ref{trip_nonl_subs}) the value
$\psi^{0}_{\rm{a}}(t)$ by the current solution $\psi_{\rm{a}}(t)$
that is found from underlying equations
(\ref{trip_dina},\ref{trip_dine},\ref{trip_ding1},\ref{trip_ding2}).
Actually, the substitution (\ref{trip_nonl_subs}) means we are
investigating the stability of the current solution instead of
that for the dark state. In fact, figure \ref{Fig_tr_non_eigvs}(b)
shows that the real parts of the eigenvalues evolve almost
identically. The only difference is that the splitting of real
parts for approximate eigenvalues is slightly delayed. Such a
coincidence shows that the current solution evolves in the close
neighbourhood to the motion of the parametrized dark state
(\ref{a0},\ref{g1},\ref{g2}). It is also to be noted that the
magnitude of $|\dot{\psi}_{\rme}|$ is always small, and the
excited state remains almost unpopulated as we can see in figure
\ref{Fig_tr_non_eigvs}(a).

It is noteworthy that a related approach was used in
\cite{cheng06}, where a feedback control scheme was presented that
designs time-dependent laser-detuning frequency to suppress
possible dynamical instability in coupled free-quasibound-bound
atom-molecule condensate systems. It was proposed to perform a
substitution analogous to (\ref{trip_nonl_subs}) which was used
for solving the control problem. On the other hand in our work
this substitution was made for the stability analysis of the dark
state.

It is also important to note that in the lambda and tripod
systems, we have phenomenologically included the loss coefficient
$\gamma$. This was done by making the one-photon detuning to be a
complex number, i.e. by replacing $\Delta \rightarrow \Delta +
\rmi\gamma$. Here $\Delta$ is again a one-photon detuning, and
$\gamma$ determines the losses. In our work, we have considered
the cases where $\Delta=0$ and $\gamma > 0$, i.e. the one-photon
resonances with losses. We stress that the presence of non-zero
losses $\gamma$ makes the adiabatic reduction easier to implement.
The losses cause the appearance of two negative real parts of
eigenvalues of the corresponding Jacobians. On the other hand, it
was shown that the losses decrease the transfer efficiency
\cite{vitan97a}, which decreases exponentially with the (small)
decay rate. However the range of decay rates, over which the
transfer efficiency remains high, appears to be proportional to
the squared pulse area. Hence, by choosing high pulse areas one
may preserve the high transfer efficiency.

Another question is a possible presence of the one-photon detuning
in the considered processes. As it was shown in section \ref{sec:6},
the relatively small one-photon detuning does not alter our
conclusions about the reduction of dimension in the three- and
four-level systems considered here. This happens if the Rabi
frequencies are large compared to the one-photon detuning and loss
rates.


\ack

The authors acknowledge the support by the EU FP7 project STREP
NAMEQUAM.



\section*{References}



\providecommand{\newblock}{}

\end{document}